\documentclass{article}

\usepackage{amsfonts,amssymb,amsthm,amsmath}

\usepackage{hyperref}
\hypersetup{
	pdfencoding=auto,
    colorlinks=true,
    linkcolor=blue,
    filecolor=magenta,      
    urlcolor=cyan,
    citecolor = violet,
}

\usepackage{graphicx}

\newcommand\blfootnote[1]{%
  \begingroup
  \renewcommand\thefootnote{}\footnote{#1}%
  \addtocounter{footnote}{-1}%
  \endgroup
}

\usepackage[margin=1in]{geometry}

\usepackage[export]{adjustbox}
\usepackage{algpseudocode}

\newtheorem{thm}{\bf Theorem}[section]

\newtheorem{lemma}[thm]{\bf Lemma}

\newcommand{\vC}{\textrm{\v{C}}}

\newcommand{\R}{\mathbb R}

\newcommand{\Z}{\mathbb Z}

\newcommand{\inv}{^{-1}}
 \newcommand{\e}{\varepsilon}

\renewcommand{\phi}{\varphi}

\newcommand{\Cech}{\v{C}ech }

\newcommand{\HP}{\HP \textrm{P}}

\newcommand{\dgm}{\mathrm{dgm}}

\usepackage[colorinlistoftodos,textsize = tiny,prependcaption]{todonotes}

\usepackage{enumitem}

\begin{document}

\title{Topological Data Analysis for True Step Detection in Piecewise Constant Signals}

\author{
Firas A.~Khasawneh\footnote{Dept.~of Mechanical Engineering, Michigan State University. \url{khasawn3@egr.msu.edu}}\\ 
Elizabeth Munch\footnote{Dept.~of Computational Mathematics Science and Engineering;
and Dept.~of Mathematics, Michigan State University. \url{muncheli@egr.msu.edu}}
}

\maketitle
\begin{abstract}  
This paper introduces a simple yet powerful approach based on topological data analysis (TDA) for detecting the true steps in a piecewise constant (PWC) signal. 
The signal is a two-state square wave with randomly varying in-between-pulse spacing, and subject to spurious steps at the rising or falling edges which we refer to as digital ringing. 
We use persistence homology to derive mathematical guarantees for the resulting change detection which enables accurate identification and counting of the true pulses. 
The approach is described and tested using both synthetic and experimental data obtained using an engine lathe instrumented with a laser tachometer. 
The described algorithm enables the accurate calculation of the spindle speed with the appropriate error bounds. 
The results of the described approach are compared to the frequency domain approach via Fourier transform. 
It is found that both our approach and the Fourier analysis yield comparable results for numerical and experimental pulses with regular spacing and digital ringing.
However, the described approach significantly outperforms Fourier analysis when the spacing between the peaks is varied.
We also generalize the approach to higher dimensional PWC signals, although utilizing this extension remains an interesting question for future research.

\blfootnote{This material is based upon work supported by the National Science Foundation under Grant Nos.~CMMI-1759823 and DMS-1759824 with PI FAK, 
and CMMI-1800466 and DMS-1800446 with PI EM. 
}
\end{abstract}


\section{Introduction}
\begin{figure}[tbp]
	\centering
	\includegraphics[width=0.95\textwidth, draft = False]{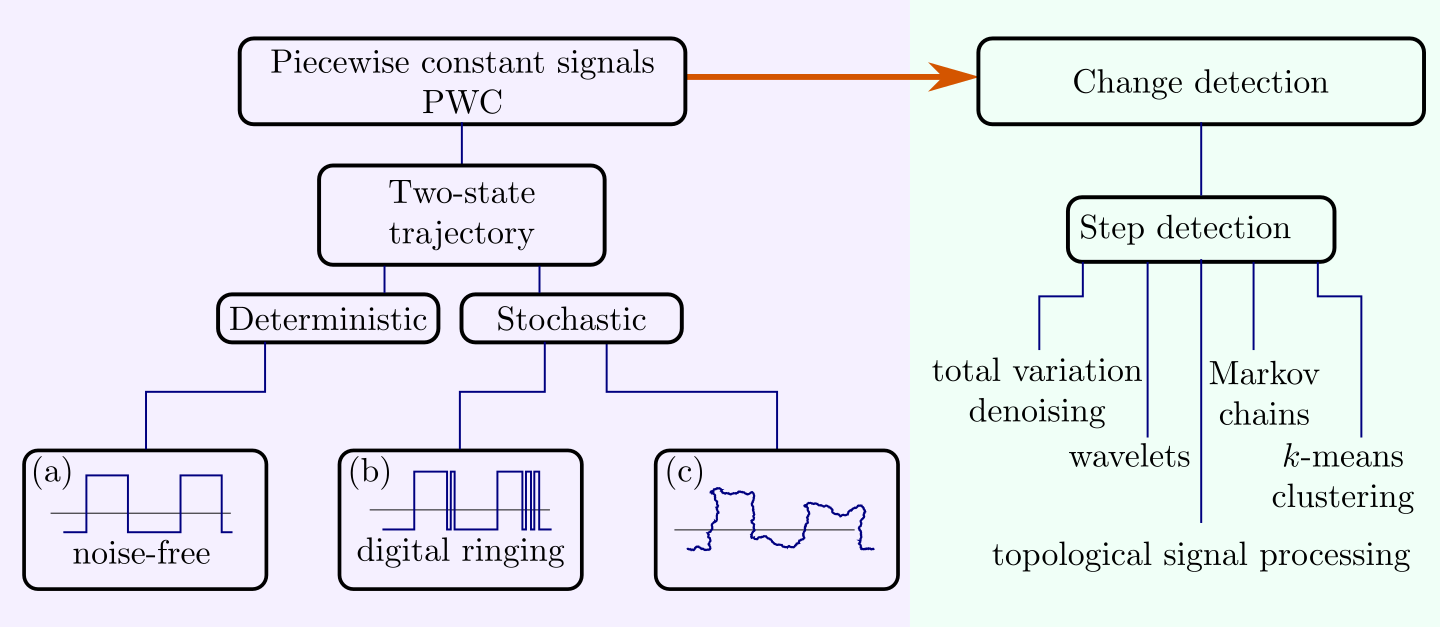}
	\caption{An overview of the categorization (left column) and analysis (right column) of the type of signals considered in this study. The paper investigates piecewise constant signals with two-state trajectories and random digital ringing (panel b in the left figure).}
	\label{fig:methodsOverview}
\end{figure}

Piecewise constant (PWC) signals are an important subclass of piecewise continuous data which occur in a variety of applications such as bioinformatics, astrophysics, geophysics, molecular biosciences and digital imagery \cite{Sowa2005, Little2010,Little2011,Little2011a,Nirody2017}. 
Figure~\ref{fig:methodsOverview}a--c shows several forms of PWC signals both deterministic and stochastic. 
The PWC waveform also occurs in the output of several sensors, which include tachometers. 
Sensors with PWC waveform are widely used and implemented in applications that involve rotating machinery such as investigating pressure fluctuations in bearings \cite{Youssef2017}, flat plate motions \cite{Jin2016}, analyzing the powertrain vibrations of vehicles \cite{Albers2016}, and for the experimental investigations of the dynamic performance of wind turbines \cite{Rahman2016}.
One important example of PWC waveform sensors are laser tachometers which are often used in machining process applications, including monitoring rotation velocity in the presence of spindle speed modulation \cite{Urbikain2016} or vibration-assisted cutting in turning \cite{Amini2017}, defect detection for friction stir welding \cite{Das2016}, measuring the feed error in tapping processes \cite{Wan2017}, and for stroboscopic sampling of vibrations in milling \cite{Honeycutt2017}. 

The output of laser tachometers assumes two logical levels low/high or on/off based on the reflectivity or contrast of the target material when subjected to a laser source. 
These tachometers often output transistor-transistor-logic (TTL) pulses triggered by the change in the reflected laser beam. 
The resulting two-state trajectory can be collected using digital or analog channels. 
If analog channels are used to collect the laser tachometer signal, then noise will be superimposed on the pulse train.  
Generally, utilizing linear filtering for noise removal from PWC signals is inefficient because both the noise and the PWC signal have a broad band frequency spectrum. 
However, since for laser tachometers the signal to noise ratio is typically large the underlying digital signal can be recovered by hard-thresholding an adequately sampled time series. 
However, hard-thresholding is not suitable in general for stochastic PWC signals such as the case shown in Fig.~\ref{fig:methodsOverview}c. 

Assuming that a representative, noise-free sample of the laser tachometer signal has been recovered, another challenge is the phenomenon where spurious pulses appear in the signal near the transitions of reflective/non-reflective target, as shown in Fig.~\ref{fig:methodsOverview}b.
We call these peaks \textit{digital ringing}.
We observed these peaks while collecting laser tachometer signals during a turning experiment. 
These spurious peaks can occur due to (1) chips or cooling fluid interrupting the laser beam in manufacturing applications, (2) vibrations of the laser tachometer holder near regions on the target with different contrast, (3) vibration of the target away or towards the laser tachometer, or (4) unintended reflections of the laser beam on non-target surfaces due to the motion of the machine.  
These spurious peaks can occur at the rising or falling edges but they do not occur as a repetitive pattern, and they may not occur at some peaks or at all combinations of cutting parameters. 
Hence, they are considered a random variable and are grouped under the umbrella of stochastic systems in Fig.~\ref{fig:methodsOverview}. 
The challenge in obtaining the once-per-revolution pulse or in counting the number of pulses per unit time is exacerbated by these spurious peaks. 
Moreover, significant complications are introduced if the spacing between the pulses is further randomly modulated either due to external disturbances or due to some noise superimposed on intentional, regular variations of the rotational speed.
In both these scenarios the interest is in finding either (a) the rising edge, (b) the peak, or (c) the falling edge where a true transition in the reflectivity of the target occurs. 
In tachometer signals it is common to detect falling or rising edges, so in this study we choose to investigate the rising edges of true transitions. 

While finding true peaks from a signal contaminated with noise may seem like an easy problem, 
the existence of spurious peaks precludes using traditional, step detection methods for peak detection and counting. 
For example, peak-finding algorithms are not useful here because true peaks can be incorrectly counted multiple times thus giving false counts and consequently yielding exaggerated spindle speeds.
In addition, setting width thresholds in the peaks algorithm requires the user to have a priori knowledge of the existing noise, which may not be feasible or practical. 
The peak finding algorithm also catastrophically fails if the spacing between peaks randomly varies due, for example, to randomness in the spindle drive or variations in the cutting load of machine tools. 
Further, since the signal is corrupted by digital ringing, typical filtering techniques for PWC signals \cite{Little2011,Little2011a} such as total variation denoising, hidden Markov chains, and wavelets are ineffective. 
Traditional clustering methods such as $k$-means clustering require knowing the number of clusters desired in advance, and thus are not useful for our need to automatically count the number of pulses.

It is worth mentioning that identifying true peaks in the presence of digital ringing as described in this 
paper is a task where humans outperform traditional computer algorithms. 
However, while humans can often identify true peaks, counting them is prone to error particularly when the number of pulses is large per unit time. 
Moreover, it is impractical to rely on human interpretation of laser tachometer signals especially in high-speed or real-time operations.

Therefore, the objective of this paper is to introduce a simple yet powerful approach for automatically detecting the true steps in a PWC two-state square wave subject to digital ringing. 
The specific application is obtaining the spindle speed of a machine tool with a laser tachometer whose signal is corrupted by spurious on/off pulses. 
We also extend the approach to the case where the spacing between the pulses varies randomly, and we discuss and extension of the theory to higher dimensional analogs, and 
we derive mathematical guarantees for the resulting change detection which enables accurate identification and counting of the true pulses.
The approach we utilize is based on Topological Data Analysis (TDA) \cite{Ghrist2014,Edelsbrunner2010,Kaczynski2004,Zomorodi2005,Ghrist2008,Carlsson2009} and it is outlined in Section \ref{sec:method}.

The prevalence of Topological Data Analysis (TDA)  has exploded in recent years due to its use in many disparate domains.
Arguably the most prevalent application is that of TDA to time series analysis and signal processing, having spawned some of the earliest results in the field \cite{Robins1998,Robins2000a,Robins2004}.
This subfield of TDA is often referred to as Topological Signal Processing (TSP) \cite{Robinson2014}.
From the insights gained from the beginnings of TSP came one of the most prominent tools in TDA, namely persistent homology \cite{Edelsbrunner2002,Zomorodian2004}, which quantifies shape and structure in data. 

Now, TSP has become a mature field in its own right.
Much of the work stems from utilizing persistent homology in conjunction with delay coordinate embeddings to give topological quantification of attractors of dynamical systems \cite{Garland2016,MacPherson2012,Alexander2015,Maletic2016,Pereira2015}.
Recent work has come in the form of classification and quantification of periodicity and quasi-periodicity \cite{Perea2016,Perea2014,Tralie2017,Robinson2015,deSilva2012}.
This work has appeared in a diverse array of applications including 
wheeze detection \cite{Emrani2014},
computer performance \cite{Alexander2012a},
market prices \cite{Gidea2017,Gidea2017a},
sonar \cite{Robinson2013c},
image processing \cite{Vejdemo-Johansson2015},
ice core analysis \cite{Berwald2014a},
machining dynamics \cite{Khasawneh2015,Khasawneh2014,Khasawneh2014a,Khasawneh2017},
and gene expression \cite{Deckard2013,Perea2015}.

In this paper, we will utilize one of the simplest examples of persistent homology, namely 0-dimensional persistence defined on points in $\R$.
Viewing our data in this way gives access to the powerful theory built up for understanding persistent homology, particularly with respect to noise.
In particular, the existence of a metric on the space of persistence diagrams leads to the powerful notion of stability \cite{Cohen-Steiner2007,Cohen-Steiner2010}.
This knowledge in conjunction with our assumptions on noise is the theoretical basis for the algorithm we develop in Section \ref{sec:method}.
While there is a growing collection of ever faster code for computation of  persistent homology \cite{Otter2015}, our restricted setting also gives rise to simplified algorithms, thus making the analysis quite fast.
An additional perk of this viewpoint is that it also allows for generalization to higher dimensional problems as outlined in Section \ref{sec:Conclusions}\ref{ssec:higherDim}.

The approach is described and tested using both synthetic and experimental data. 
The experimental apparatus includes an engine lathe instrumented with a laser tachometer which detects the change in reflectivity of a tape adhered to the circumference of the spindle. 
The resulting mean spindle speed is then reported with appropriate error bounds. 
We also compare the results to the output of Fourier analysis. 
It is found that the described approach provides comparable results to Fourier for regular pulse waves with digital ringing; however, the numerical calculations show that the described approach outperforms Fourier analysis when the spacing between the pulses is varied.
%


\section{Background}
This section provides the necessary background for motivating and presenting the new approach for pulse counting using persistence diagrams. 
Section \ref{sec:current_methods} discusses existing methods for pulse counting and spindle speed calculations. Sections \ref{sec:0dim_persistent}--\ref{sec:MST} present the background theory on persistent homology and describes how it relates to the current work.

\subsection{Current methods for counting and rotational speed calculation} \label{sec:current_methods}
Sensors for detecting rotary motion include proximity sensors, photoelectric sensors, and encoders. 
These sensors output pulses that can be counted and used to find the speed of the shaft, typically, in units of revolutions per minute (RPM). 
The pulse signals are piecewise continuous (PWC) functions with two logic levels: high or on, and low or off. 
The quality of the output data depends on the number of pulses per revolution which affects the data resolution, as well as the symmetry of the pulses, which influences the accuracy and consistency of the data. 

Once the pulses are obtained, there are generally two techniques for determining the corresponding RPM: 1) frequency measurement approach (calculate RPM from pulse count and pulse frequency), and 2) period measurement approach (calculate RPM from pulse count and pulse period). 
The frequency approach involves transforming the signal into the frequency domain using a Fourier transform; however, the Fourier transform of a PWC signals can have slow convergence \cite{Stephane2009}.
Other counting algorithms include local maxima or peak detection with wavelet transforms or other methods \cite{Little2011}. 
However, the interest in this study is in detecting true peaks, not all peaks. 
Therefore, utilizing conventional methods for peak detection must be combined with a threshold for rejecting false peaks or retaining true ones. 
It may be tempting to utilize a statistical measure such as the variance or standard deviation of the pulse duration. 
However, since the data can vary from bimodal to uniform depending on the amount of noise introduced, statistical dispersion measures are generally not effective. 

When there is noise superimposed on the signal (for example due to  collecting the signal on an analog channel) this noise can be removed by hard thresholding. 
If the noise component is so large that the pulse structure can no longer be distinguished as shown in Fig.~\ref{fig:methodsOverview}c, the denoising becomes more difficult because the traditional approach of low-pass filtering typically introduces large spurious oscillations in PWC signals \cite{Stephane2009}. 
In this case, either the viewpoint of piecewise constant smoothing, or that of level-set recovery can be used for determining the location of the jumps \cite{Little2011}. 
In this paper, we focus on signals where the small amount of noise superimposed on the pulse amplitude can be removed by thresholding, but where spurious random pulses occur at the rising or falling edges and where the spacing between peaks can randomly vary. 
This type of noise cannot be removed by thresholding, or by traditional filtering techniques. 
Varying the period length between two consecutive pulses significantly complicates the analysis, and without a proper method for change detection can lead to poor RPM calculations.
Therefore, there is still a need for new, robust tools for pulse detection in PWC signals and this paper presents a method for reliable pulse-counting using 0-dimensional persistent homology.

\subsection{0-dimensional Persistent homology} 
\label{sec:0dim_persistent}
Persistent homology \cite{Robins1998,Robins2000a,Edelsbrunner2002,Zomorodian2004}, a tool arising from Topological Data Analysis \cite{Carlsson2009,Ghrist2008}, seeks to quantify shape and structure in data sets.
Algebraic topology \cite{Hatcher,Munkres2000} is a field of mathematics which quantifies qualitative similarities in the structure of spaces.
One such method for quantification is homology, which, given a topological space, provides a vector space\footnote{We work with field coefficients, most typically $\Z_2$, so the homology group is, in fact, a vector space.} for each dimensional structure being studied.
This paper will only focus on dimension 0, which quantifies connected components.

The intuition behind persistent homology for a point cloud data set is to increase a connectivity parameter, and quantify how the topological structure changes.
This powerful method can find interesting, higher dimensional structure, using each dimension of homology, however, we will look at the simplest version for the purposes of our problem.
Zero-dimensional persistent homology  quantifies how the clusters change when viewed at different scales.  
In fact, 0-dimensional persistent homology is closely related to classical clustering methods such as single-linkage hierarchical clustering, dendrograms, and minimal spanning trees \cite{Murtagh2011}.
Here, we adopt the view of the procedure as a restricted case of persistent homology in order to gain understanding and predictive power in our analysis, thus resulting in Thm.~\ref{thm:MainResult}.  

In the general setting, assume we are given a point cloud $\chi \subset \R^D$ with $|\chi| = n$.  
We can define a function $f_\chi:\R^D \to \R$ by 
$f_\chi(x) = 2\|x - \chi\|$ where $\|x-A\| = \inf_{y \in A} \|x-a\|$ for any set $A \subset \R^D$.
The set of points for which $f_\chi(x) \leq r$ is the union of $D$-dimensional balls of radius $\frac{1}{2}r$ centered at the points of $\chi$;
we write this as $f_\chi\inv(-\infty,r]$.
If we allow $r$ to increase from $0$, initially, $f_\chi\inv(-\infty,r]$ has $n$ distinct connected components.
However, as $r$ is increased, these components will merge together until we are finally left with only one connected component.
In particular, these mergings happen at the instant two disks touch, and thus when $r$ is equal to the distance between the associated points.  
We say that a connected component dies when it merges with another connected component; that is, a death occurs any time two clusters merge.
We can keep track of the function values at which these deaths occur in the following manner.

A function value $r$ is a homological critical value if the number of connected components decreases at function value $r$.
The multiplicity of a critical value is the net decrease in the number of connected components.
We define $\dgm(\chi) \subset \R$ to be the collection of homological critical values, with number of copies equal to the multiplicity.
The resulting set of values is called a 0-dimensional persistence diagram, notated $\dgm = \{d_1 \leq d_2 \leq \cdots \leq d_k\}$.%
\footnote{Experts will notice that, in general, persistence diagrams are given as collections of points in $\R^2$.  Since for 0-dimensional persistence on point clouds, all components are born at 0, the true persistence diagram would be the collection $\{(0,d_1),\cdots,(0,d_k)\}$.  We elect to drop the repeated first coordinate in order to simplify the description.}
One observation that will be useful for interpretation is that the number of connected components of $f\inv(-\infty,r]$ is one more than the number of $d_i \in \dgm$ with $d_i > r$.

We will also use the slightly more general formulation of persistence for any subset $A \subset \R^D$.
If $f_A: \R^D \to \R$ is given by $f_A(x) = 2\|x-A\|$, $\dgm(A)$ is the collection of function values for which the number of connected components of $f\inv(-\infty,r]$ changes, again with multiplicity.
In order to ensure that the persistence diagrams are finite, we assume that $A$ has finitely many connected components.

\begin{figure}[tb]
	\centering
	\includegraphics[height = 1.3in,draft = False,valign = t]{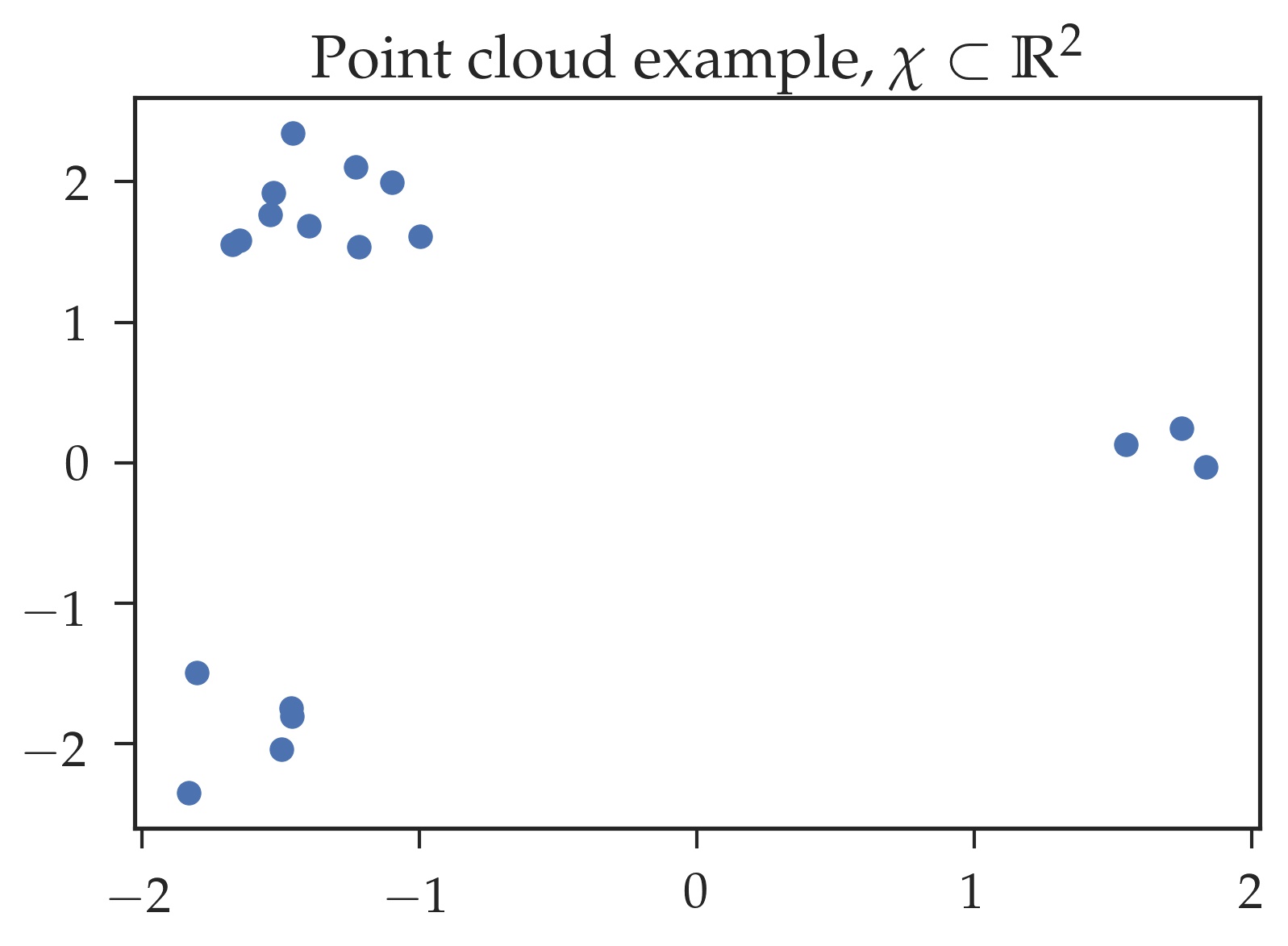}
	\includegraphics[height = 1.35in,draft = False,valign = t]{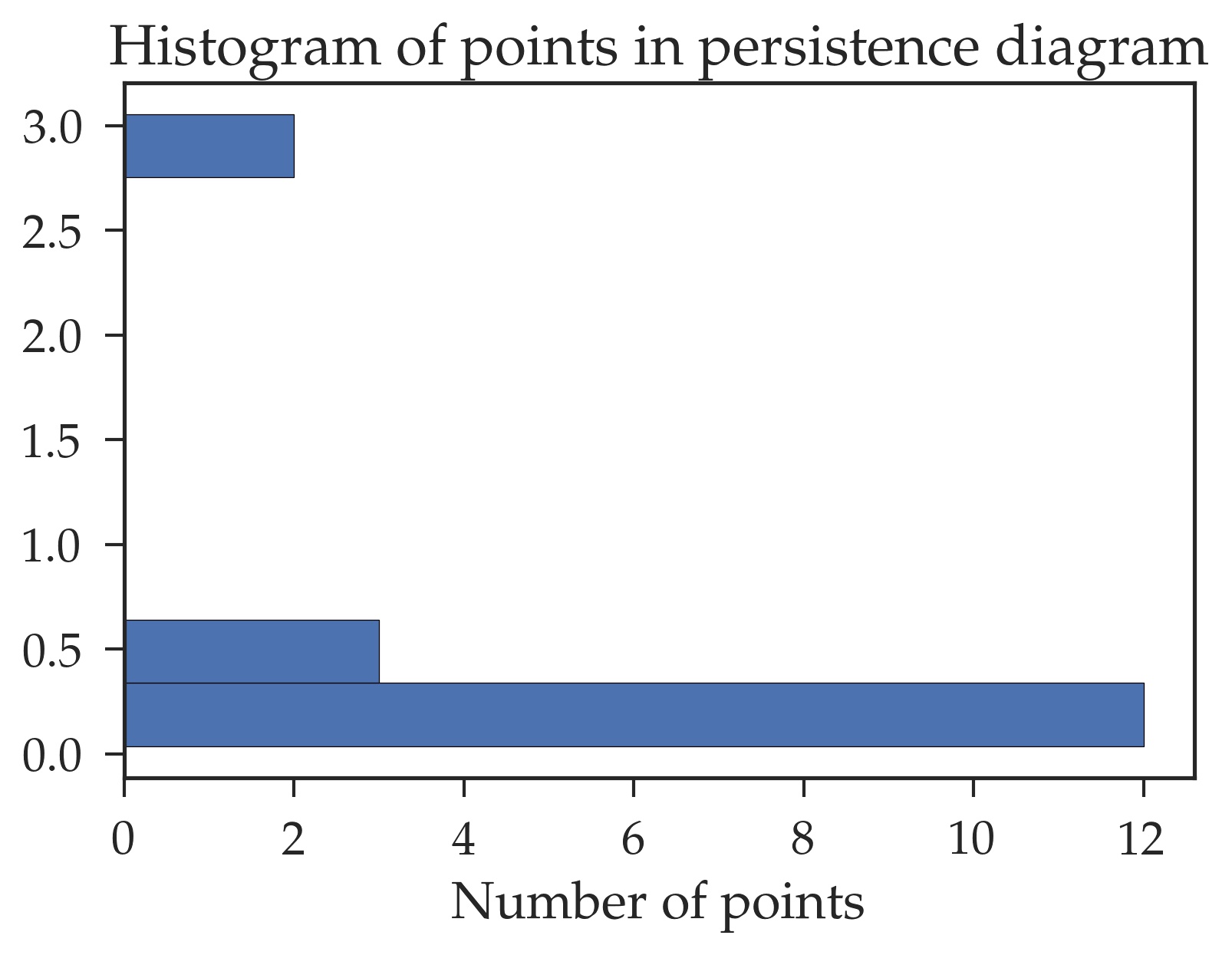}
	\includegraphics[width = .2\textwidth, draft = False,valign = t]{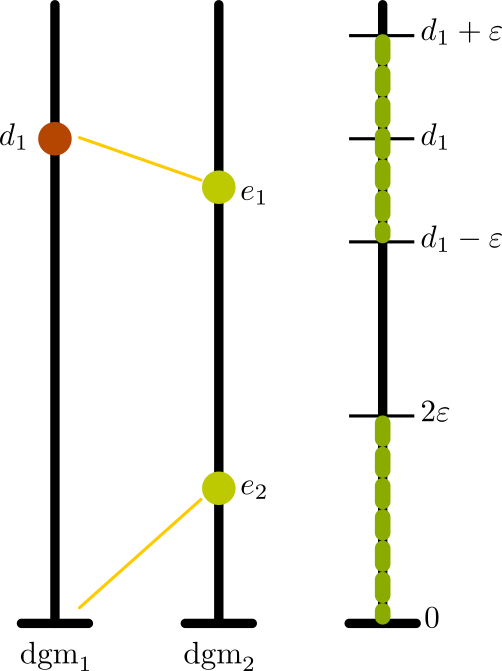}
	\caption{Left, an example point cloud with its 0-dimensional persistence diagram drawn as a histogram at center.  Right, an example of the pairing for two persistence diagrams used to compute the bottleneck distance.}
	\label{fig:point_cloud_cluster_hist}
\end{figure}

\subsection{Stability of Persistence Diagrams}
\label{ssec:stabilityBkgd}

The main reason for thinking of this information as a persistence diagram is that these structures come with a metric known as the bottleneck distance.
Here we describe the metric in the restricted 0-dimensional persistence diagram setting; see e.g.~\cite{Edelsbrunner2010} for the full definition.
Given two 0-dim persistence diagrams, $\dgm_1 = \{d_1 \leq d_2 \leq \cdots \leq d_k\}$ and $\dgm_2 = \{e_1 \leq e_2 \leq \cdots \leq e_\ell\}$, a partial matching $\eta$ is a bijection between subsets of the two diagrams $\eta: A \to B$, $A \subset \dgm_1$, $B \subset \dgm_2$.
The cost of a matching is defined to be 
\begin{equation*}
c(\eta) = \max\left(  
\bigg\{ |a - \eta(a)| \bigg\}_{a \in A} \cup
\bigg\{ \frac{a}{2} \bigg\}_{a \in \dgm_1 \setminus A} \cup   
\bigg\{ \frac{b}{2} \bigg\}_{b \in \dgm_2 \setminus B}   
\right)
\end{equation*}
and the bottleneck distance 
$d_B(\dgm_1,\dgm_2) = \min_{\eta} c(\eta)$
is the minimum cost of the possible matchings.
Notice that because we are restricting our diagrams to finitely many points, the set of matchings is finite, thus this minimum is always achieved.
In addition, it is possible that multiple matchings achieve the minimum, so we will reference such a matching as \textit{a} min-cost matching. 

Consider, for example, a diagram consisting of a single point $\dgm_1 = \{d_1\}$ vs a diagram with two copies of the same point $\dgm_2 = \{e_1,e_2\}$.
WLOG, assume $e_1\geq e_2$.
The only possible matchings are where $\eta_1(d_1) = e_1$, $\eta_2(d_1) = e_2$, or $\eta_3$ which matches nothing.  
The scores of these are
$c(\eta_0) = \max\left\{\tfrac{d_1}{2}, \tfrac{e_1}{2}, \tfrac{e_2}{2}\right\}$, 
$c(\eta_1) = \max\left\{|d_1-e_1|, \tfrac{e_2}{2}\right\}$,
and
$c(\eta_2) = \max\left\{|d_1-e_2|, \tfrac{e_1}{2}\right\}$.
In the example of the two persistence diagrams in the right of Figure \ref{fig:point_cloud_cluster_hist}, the score is lowest using $\eta_1$, so the distance between the diagrams is $\max\left\{|d_1-e_1|, \tfrac{e_2}{2}\right\}$.

The bottleneck distance is particularly useful due to the stability theorem \cite{Cohen-Steiner2007}.
Recall that the Hausdorff distance between sets $A,B \subset \R^D$ is 
\begin{equation*}
d_H(A,B) = \max \left\{ \sup_{a \in A} \|a-B\|, \sup_{b \in B} \|b-B\| \right\}.
\end{equation*}
Then the stability theorem is as follows.
\begin{thm}
[\cite{Cohen-Steiner2007}]
\label{thm:stability}
Under mild assumptions on the sets $A,B \subset \R^D$\footnote{The distance functions $f_A$ and $f_B$ must be what is known as \textit{tame} in the TDA literature.  
As all sets considered in this paper have finitely many connected components in $\R$, we will always satisfy this assumption. }, 
\begin{equation*}
d_B(\dgm(A), \dgm(B)) \leq d_H(A,B).
\end{equation*}
\end{thm}
This theorem is particularly useful when we take the view, as we will need in Section \ref{sec:method}, that $B$ is a noisy point cloud approximation of $A$. 
In this case, $d_H(A,B)$ is small, so the resulting persistence diagrams will be close as well.

Let us now explore what it means for a diagram to be close to another diagram.
Assume we again have a diagram with a single point $\dgm_1 = \{d_1\}$.
For $\e$ small relative to $d_1$, we can think of the set of diagrams within distance $\e$ of $\dgm_1 = \{d_1\}$.
Say the second diagram is $\dgm_2 = \{e_1,\cdots,e_n\}$, so the options for matchings are $\eta_i$ matching $d_1$ to $e_i$ for $i=1,\cdots,n$, and $\eta_0$  matching nothing.
Assume $d_B(\dgm_1,\dgm_2) \leq \e$ and let $\e < d_1/3$.  
If a minimum cost matching is $\eta_i$, this implies $|d_1 - e_i| \leq \e$, and $e_j \leq 2\e$ for all $j \neq i$.
This means that there must be exactly one point, $e_j$, within distance $\e$ of $d_1$, and all remaining points are in $[0,2\e]$.
As shown in the rightmost portion of Figure \ref{fig:point_cloud_cluster_hist}, this means there is exactly one point in the top green region, and any remaining points are in the bottom green region.

What will be useful in Section \ref{sec:method}\ref{ssec:diamOfPulses} is the set of diagrams close to the diagram which has $n$ copies of the point $d_1$, e.g.~$\dgm_n = \{d_1,d_1,\cdots,d_1 \}$, .
In this case, any diagram within bottleneck distance $\e$ (again sufficiently small $\e < d_1/3$), will have exactly $n$ points in $[d_1-\e,d_1+\e]$, with all remaining points less than $2\e$.

\subsection{Computation of the persistence diagram using MSTs}
\label{sec:MST}
In order to do computations, we convert this information into combinatorial structures.
Following \cite{Robins1998,Robins2000a}, we compute persistence using a minimal spanning tree.
A \textit{graph} $G = (V(G),E(G))$ consists of a finite list of vertices $V(G)$ and a set of edges, $E(G)$, between them.
A graph is \textit{complete} if every pair of vertices has an edge between them.
A \textit{subgraph} $A \subseteq G$ is a subset of $G$ which is itself a graph.
That is, $A = (V(A),E(A))$ where $V(A) \subseteq V(G)$, $E(A) \subseteq E(G)$.

A \textit{path} in $G$ is a sequence of vertices $v_0,\cdots,v_n$ such that there is an edge between every adjacent pair: $(v_i,v_{i+1}) \in E(G)$ $\forall i$.
A path is a \textit{cycle} if $v_0 = v_n$.
A graph is \textit{connected} if there is a path between every pair of vertices.
A graph is a \textit{tree} if it is connected and there are no cycles.
A subgraph $T \subseteq G$ which is a tree and with $V(T) = V(G)$ is called a \textit{spanning tree}.

A weighted graph $G = (V(G),E(G),\omega)$ is a graph with a real value, called a weight, associated to each edge: $\omega: E(G) \to \R$.  
The \textit{total weight} of a spanning tree of a weighted graph is the sum of the weights on the edges.
A spanning tree is called a \textit{minimal spanning tree (MST)} if it has minimal total weight amongst the set of spanning trees.
If $G$ is not connected, no MST exists as there is no connected subgraph using all the vertices.
A \textit{minimal spanning forest (MSF)} is an acyclic subgraph such that the restriction to each connected component of $G$ is a MST.

Given a finite set of points $\chi \subset \R^D$, the \Cech graph for parameter $r>0$ is the complete graph $\vC(\chi,r)$ with vertex set in 1-1 correspondence to the points in $\chi$, with all edges $(x,y)$ for which $\|x-y\| \leq r$, and with edge weight equal to the distance between the associated points.
First, notice that if $r\leq s$, $\vC(\chi,r) \subseteq \vC(\chi,s)$.
Second, if $r$ is larger than the diameter of $\chi$ (which is finite since $\chi$ is finite), $\vC(\chi,r)$ is a complete graph and $\vC(\chi,r) = \vC(\chi,s)$ for all $s \geq r$.
Thus, we denote this graph as $\vC(\chi,\infty)$.

Our first intuition of understanding the connected components of the union of disks as described above is to watch the connected components of $\vC(\chi,r)$ change as $r$ increases.  
It is a consequence of the celebrated Nerve Lemma \cite{Hatcher} that the connected components of $\vC(\chi,r)$ match up with the connected components of the union of disks of radius $\tfrac{r}{2}$; equivalently with the connected components of $f_\chi(-\infty,r]$ in the notation of the previous section.
So, we just need to determine the list of function values for which the connected components of $\vC(\chi,r)$ change.

Let $T$ be a MST for $\vC(\chi,\infty)$\footnote{This is sometimes called a ``Euclidean spanning tree''.}.
It is an immediate consequence of the results \cite{Robins2000a} that the list of weights on the edges of $T$ is exactly $\dgm(\chi)$.
Restated in our notation, this is the following.
\begin{lemma}[{\cite[Lem.~3]{Robins2000a}}]
Let $T$ be a MST for $\vC(\chi,\infty)$,
and let $T_r$ be the restriction to the set of edges of weight at most $r$ so that $T_r \subset \vC(\chi,r]$.
Then $T_r$ is a MSF for $\vC(\chi,r]$.
\end{lemma}
Since a MSF of a graph has the same connected components as the graph itself, this means that the changes in the connected components happen exactly at the values of the weights in $T$.
Putting this all together, we have the following theorem.
\begin{thm}
For a finite set of points $\chi$, $\dgm(\chi)$ is the set of weights on the edges of a MST of $\vC(\chi,\infty)$.
\end{thm}

The final thing to note is that we will largely be interested in points in $\R$ (so $D=1$).  
Say $\chi = \{a_0,\cdots,a_k\}$.
In this case, it is obvious that the MST is exactly the graph with edges $\{(a_i,a_{i+1})\}_{i=0}^{k-1}$, and so the diagram $\dgm(\chi) = \{(a_{i+1}-a_i)\}_{i=0}^{k-1}$.

%


\section{Method}
\label{sec:method}

\subsection{Basic model assumptions}
\label{ssec:modelAssumptions}
\begin{figure}[tb]
\centering
\includegraphics[width = .23\textwidth,valign = t, draft = False]{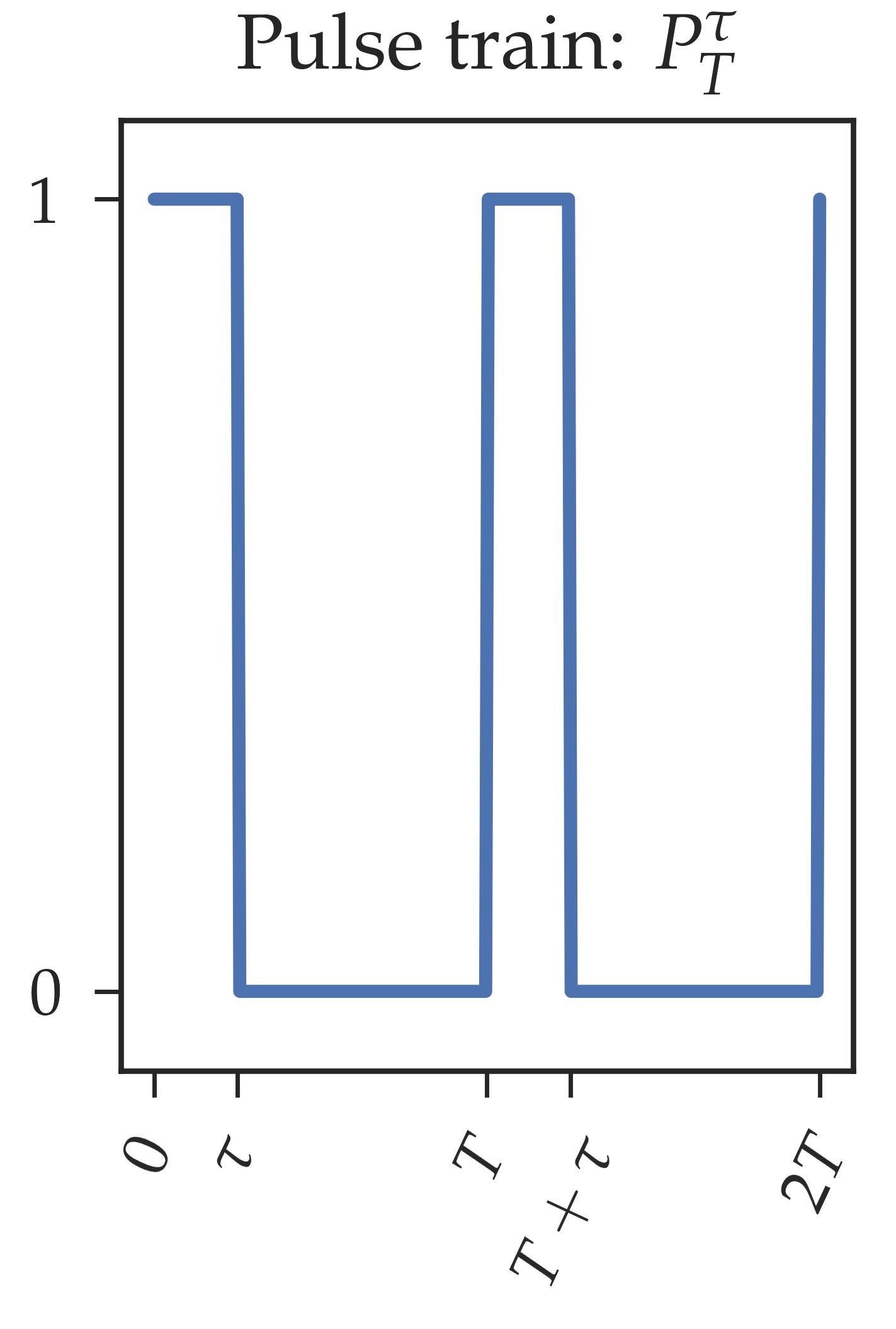}
\includegraphics[width = .36\textwidth,valign = t,, draft = False]{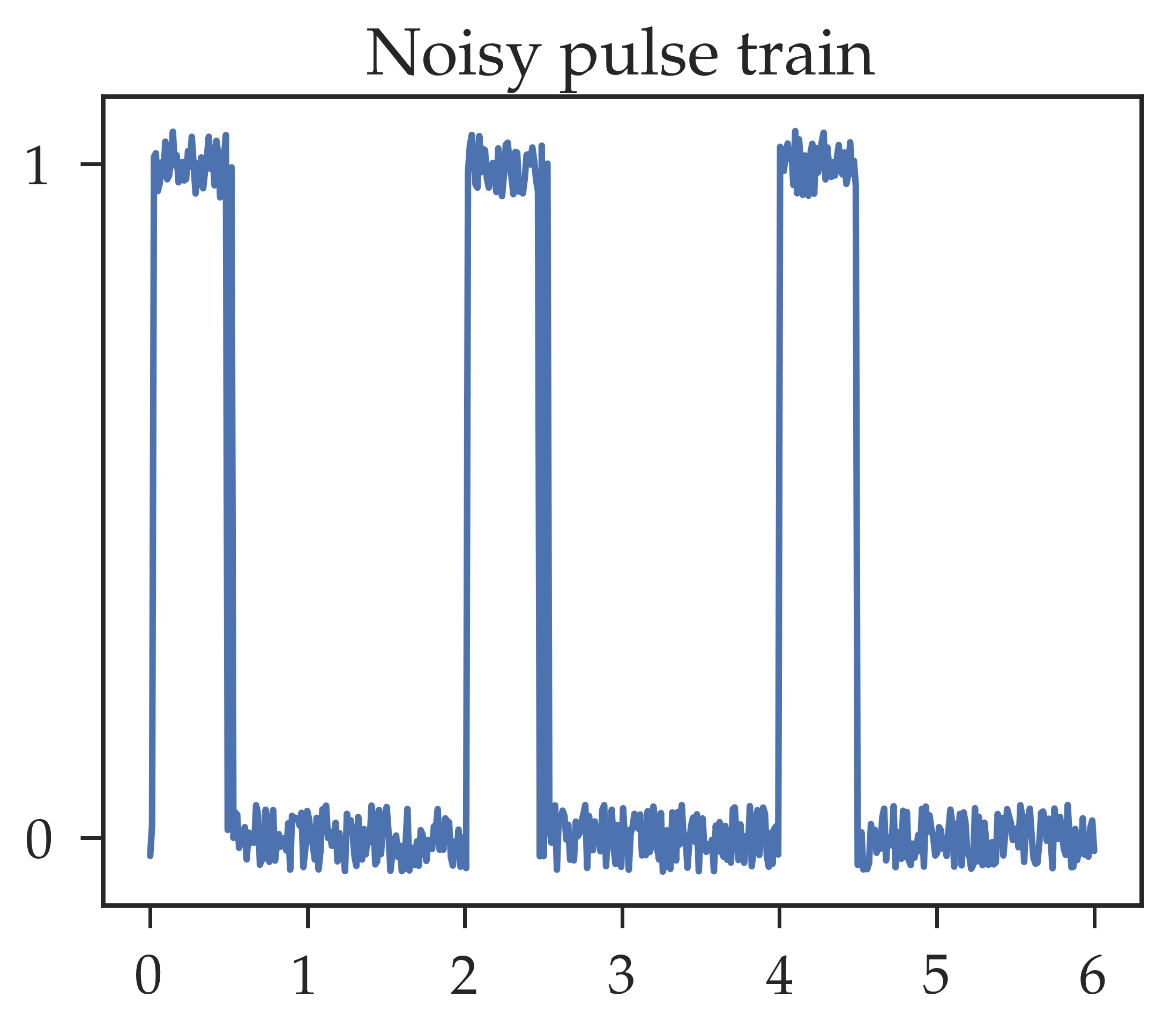}
\includegraphics[width = .36\textwidth,valign = t,, draft = False]{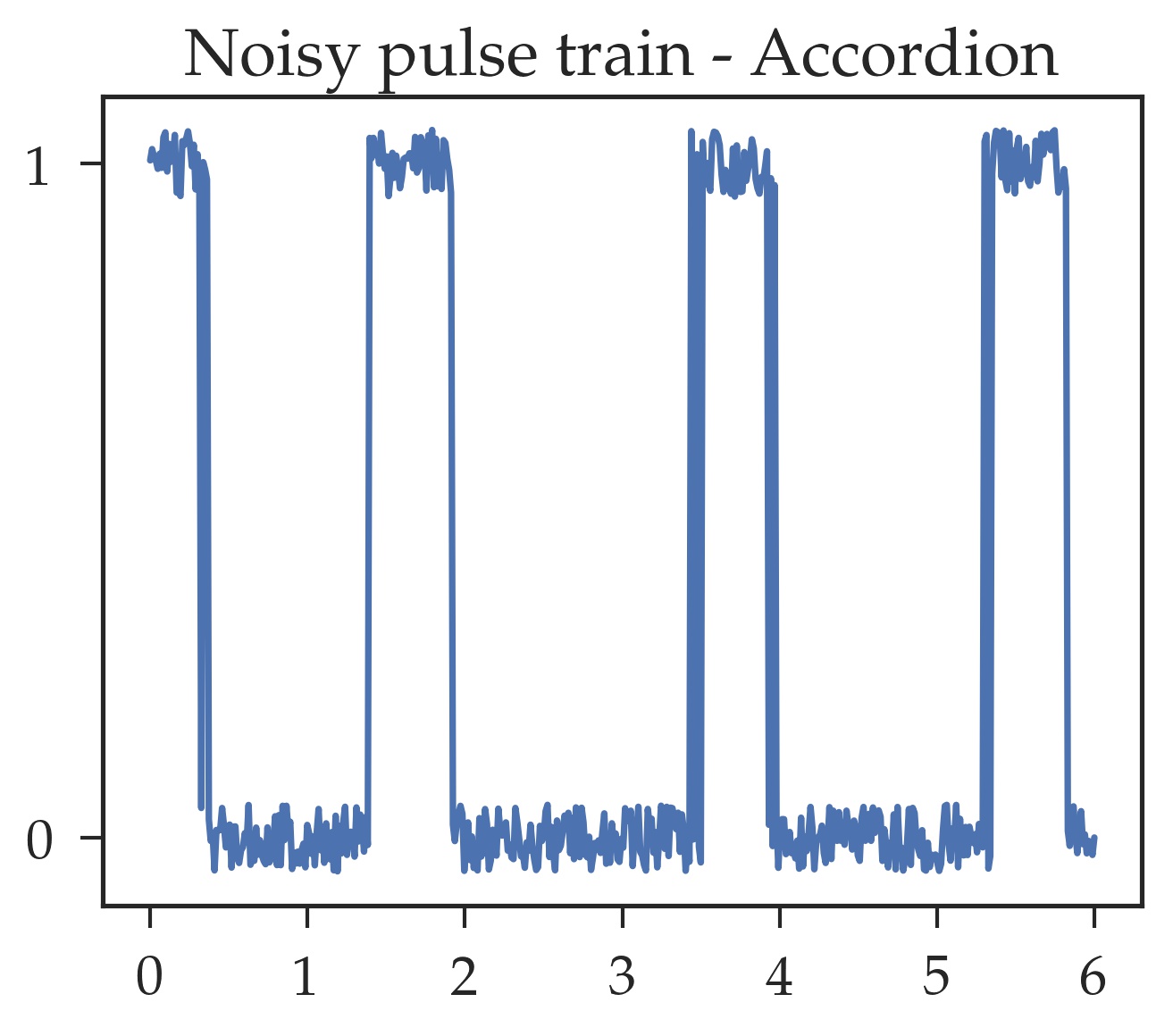}
\caption{Left, an ideal pulse train $P_T^\tau$.  
Center and right, noisy instances of the pulse train, both with 
$T = 2$, 
$\tau = .5$, 
$\alpha = .05$, and $\beta = .05$.
The center figure has $\e = 0$ and the right figure has $\e = .5$. Note the uneven, random pulse spacing for $\epsilon \neq 0$.}
\label{fig:PulseTrain}
\end{figure}

The initial data is a time series $X$ defined on $\{a=t_0 <t_1<\cdots<t_N=b\}$
given by $\{X(t_0), X(t_1),\cdots,X(t_N)\}$.
We assume that this data approximates the pulse wave $P_T^\tau$ given by
\begin{equation*}
P_T^{\tau}(t) = 
\begin{cases}
1 & \text{if } 0 \leq (t \mod T) \leq \tau\\
0 & \text{else}
\end{cases}
\end{equation*}
and shown in Figure \ref{fig:PulseTrain}.
We will assume that the duty, $\delta = \tau/T$, is less than $0.5$ since otherwise we could swap our analysis to check for the pulses at 0 instead of the pulses at 1.

We first assume that our noisy data is given by 
\begin{equation}
\label{eq:ModelSimple}
X(t) = P_T^\tau(t+\delta_x) + \delta_y
\end{equation}
where 
$\delta_x \sim \mathrm{unif}(-\alpha\cdot \tau,\alpha\cdot \tau)$ and 
$\delta_y \sim \mathrm{unif}(-\beta,\beta)$ with $\alpha \in [0,1/2]$ and $\beta \in [0,1]$.
For simplicity, we assume the noise in $y$, $\delta_y$, separates the high and low values, so  
$\beta \ll \tfrac{1}{2}$.
The noise in $x$, $\delta_x$, serves to give incorrect pulse values when we near the beginning or end of a pulse by looking forward and backward in time to reach an output.  
Since $\alpha$ is given as percentage of $\tau$, and we will not be able to see anything if $\alpha$ is large enough to entirely erode the off-cycle with digital ringing, we assume 
$\alpha \leq \tfrac{T-\tau}{3\tau}$.
For practical applications with a small $\tau$, this is satisfied for any $\alpha \in [0,1/2]$.

It will be important later to note that with these assumptions on $\beta$ and $\alpha$, the Hausdorff distance between $M$ and $\chi = \{t \mid X(t) \geq \tfrac{1}{2}\}$ for any realization of $\{X(t)\}$  dense enough in $t$ will be bounded by $\alpha \tau$.


\subsection{Counting pulses and RPM calculation using persistence}
\label{ssec:diamOfPulses}

\begin{figure}[tb]
\centering
\includegraphics[width = .5\textwidth, draft = False]{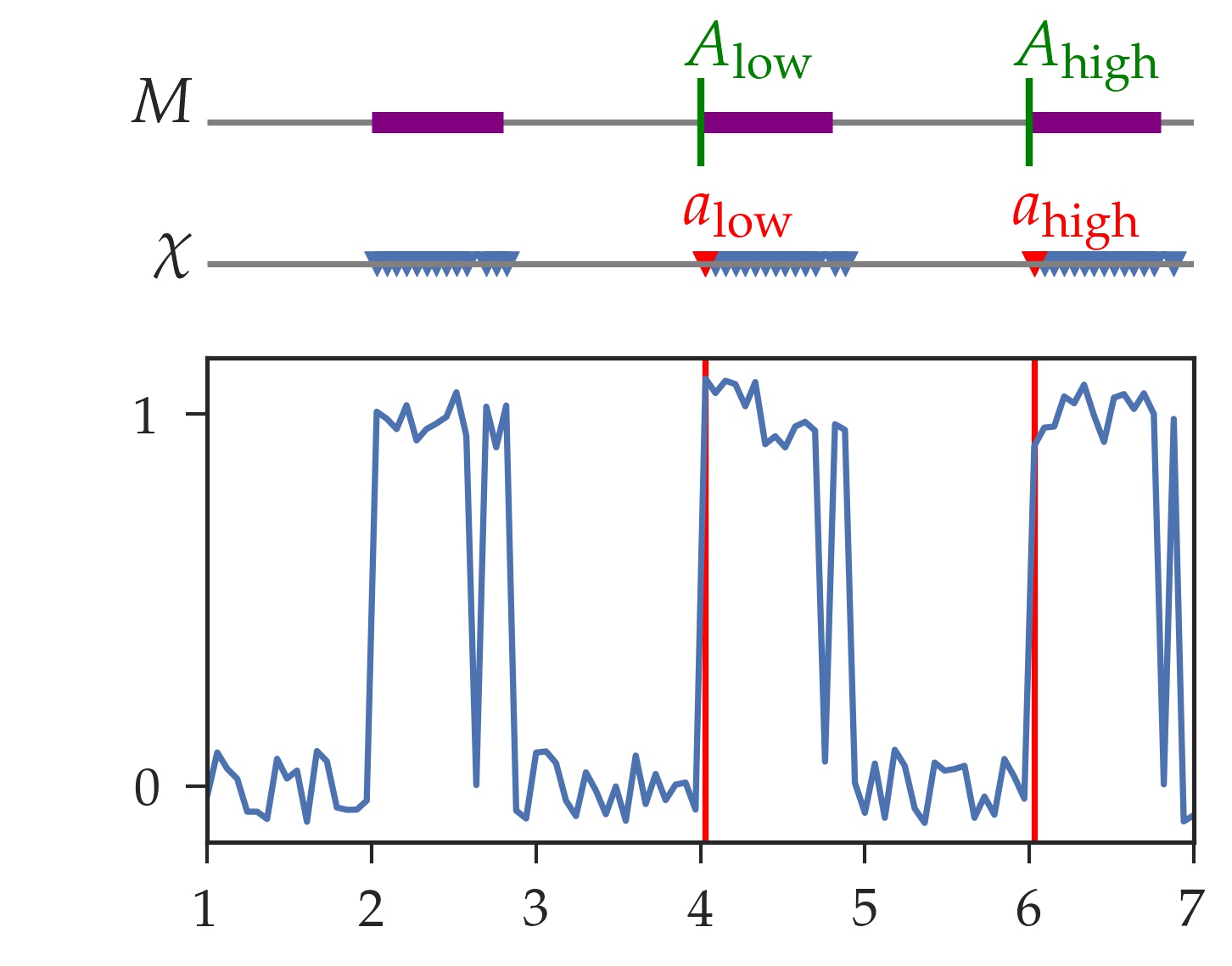}
\includegraphics[width = .3\textwidth, draft = False]{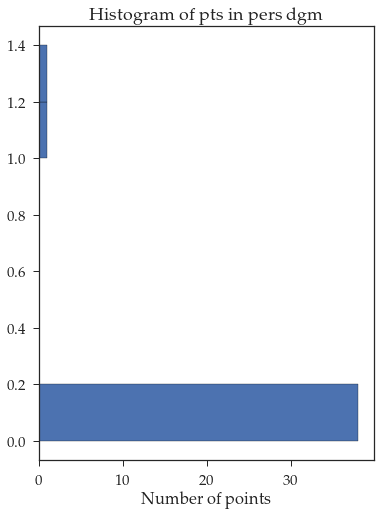}
\caption{
A pulse train, with $T = 2$, $\tau = .8$, $\alpha = .2$, $\beta = .1$, is shown at left. 
The corresponding sets $\chi$ and $M$ are drawn above. The choices of domain for the RPM computation is shown by the green lines.
The histogram for the points in the 0-dimensional diagram for the example of Figure \ref{fig:PulseTrain2}.  Note that there is a large split between the non-empty bins, so we can choose any threshold $\mu \in [0.2,1.0]$.}
\label{fig:PulseTrain2}
\end{figure}

Counting the number of pulses in a range $[A,B]$ can be thought of as counting the number of connected components of $M = (P_T^\tau)\inv(\{1\}) \cap [A,B]$.
With this in mind, we use a special case of 0-dimensional persistence to determine this from the point cloud approximation of $M$, namely
$\chi = \{t_i \mid X(t_i) >\tfrac{1}{2}\}$.
See Figure \ref{fig:PulseTrain2} for an example.

In order to determine the persistence-based RPM $\Omega_P$, we need to find the range on the $t$-axis containing the full periods, and count the number of full periods there. 
In some sense, this procedure is measuring the voids between the components of $M$, rather than measuring the components of $M$ themselves.
The following theorem essentially outlines an algorithm, which, given controlled enough noise, dense enough sampling, and enough visible pulses, calculates the number of pulses in that region using the persistence diagram $\dgm(\chi)$.

\begin{thm}
\label{thm:MainResult}
Assume a time series $X(0),X(t_1),\cdots,X(t_N)$ is data drawn from the model of Eq.~\ref{eq:ModelSimple} with $\alpha < \frac{1}{5}(\frac{T}{\tau}-1)$, and $\beta < .5$, and with times $\{A=t_0<t_1<\cdots<t_N = B\}$ evenly spaced such that $t_i-t_{i-1} < \tau(1-2\alpha)/2$ and $(B-A)/T > 3$.  
Set $\chi = \{t_i \mid X(t_i) >\tfrac{1}{2}\} = \{a_1\leq\cdots\leq a_m\}$ and denote the resulting persistence diagram $\dgm(\chi) = \{b_1\leq\cdots\leq b_\ell\}$. 
Choose $\mu   \in (d_j, d_{j+1})$ for 
$j = \mathrm{argmax}_k\{d_{k+1}-d_k\}$.
Then 
$|\{d \in \dgm(\chi) \mid d > \mu\}| - 1$
is the number of pulses in the range
\begin{equation}
\label{eq:DefnALowHigh}
A  < A_{\mathrm{low}} := 
\left( \left\lceil \frac{A-\tau}{T} \right\rceil +1\right)T 
< 
\left\lfloor \frac{B}{T} \right\rfloor T =: A_{\mathrm{high}}
\leq B.
\end{equation}


\end{thm}

In particular, since $A_{\mathrm{low}}$ and $A_{\mathrm{high}}$ can be approximated within $\alpha \tau$ by 
\begin{align*}
a_{\mathrm{low}} &= \min\{a_i \in \chi \mid a_i-a_{i-1} > \mu\}, \textrm{ and}\\
a_{\mathrm{high}} &= \max\{a_i \in \chi \mid a_i-a_{i-1} > \mu\},
\end{align*}
this theorem allows us to approximate the RPM in the following sections as 
\begin{equation}
\label{eqn:RPM}
\mathrm{\Omega_P} = 
\frac{\#\{d \in \dgm(\chi) \mid d \geq \mu\} - 1 }{a_{\mathrm{high}} - a_{\mathrm{low}}}.
\end{equation}.

\begin{proof}


First, note that if $M$ has $m$-connected components, there will be $m-1$ mergings of components that occur exactly at distance $T-\tau$.
Thus, the persistence diagram of $M$, $\dgm(M)$, has $m-1$ copies of $d_1 = T-\tau$.
By definition of the noise parameters $\alpha$ and $\beta$, we know that the Hausdorff distance between $M$ and $\chi$ is at most $\alpha \tau$.  
So, by the stability theorem, Thm.~\ref{thm:stability}, we have that
$d_B(\dgm(M),\dgm(\chi))<\alpha\tau$.

Thus, we turn our attention to computation of the 0-dimensional persistence of $\chi$.
As discussed previously, the MST for points in $\R$ has edges $\{(a_i,a_{i-1})\}_{i=0}^{k-1}$.
Hence, the set of pairwise distances  $\{a_1-a_0, a_2-a_1,\cdots, a_n-a_{n-1}\}$ is exactly the 0-dimensional persistence diagram $\dgm(\chi)$.

The set of possible 0-dimensional diagrams within bottleneck distance 
$\alpha\tau$ of $\dgm(M)$ have exactly $m-1$ points in 
$[(T-\tau )-\alpha\tau, (T-\tau) + \alpha\tau]$, 
and all remaining points are in $[0,2\alpha\tau]$.
Requiring $\alpha < \frac{1}{5}(\frac{T}{\tau}-1)$ means that there will be a distinct split in $\dgm(\chi)$; in particular that the largest difference between points in the diagram will occur between a point in $[0,2\alpha\tau]$ and a point in $[(T-\tau )-\alpha\tau, (T-\tau) + \alpha\tau]$.
These points are then notated as $d_j$ and $d_{j+1}$ where 
$j = \mathrm{argmax}_k\{d_{k+1}-d_k\}$.

We choose a threshold $\mu$ between the collections, and count $\#\{d \in \dgm(\chi) \mid d \geq \mu\} + 1$; that is to say, the number of points in the diagram in the interval $[(T-\tau )-\alpha\tau, (T-\tau) + \alpha\tau]$, so this is $m-1$.

While it is an annoying case study in number theory to determine exactly how many pulses appear in the interval $[A,B]$, what we can say is that if we cut off the first (possibly partial) pulse seen, and the last (possibly partial) pulse seen by looking in the interval $[A_{\mathrm{low}}, A_{\mathrm{high}}]$ defined in Eq.~\ref{eq:DefnALowHigh} 
then the number of pulses is $m -2$.
Thus, the theorem follows.
\end{proof}

\subsection{Counting pulses and RPM calculation using Fourier transform}
\label{sec:fourier_rpm}
The pulse counting and RPM calculation can be obtained by transforming the signal to the frequency domain. Specifically, the one-sided Fourier transform of the signal is obtained after subtracting the mean value from it to remove the DC component from the spectrum. Then, the frequency $f$ in Hertz corresponding to the first amplitude $A$ that satisfies $A > A_{\text{max}}/w$ is found, where $w$ is a scalar, and the Fourier-based RPM $\Omega_{\text{F}}$ is calculated according to
\begin{equation}
\label{eq:RPMxFourier}
\Omega_{\text{F}} = f_1 \times 60.
\end{equation}
In this study we chose $w=3$ since it provided accurate results for the studied cases with constant peak spacing. However, other values of $w$ did not improve the accuracy of the method for the cases with non-constant peak spacing, and therefore the deterioration in the performance of the Fourier-based approach  for these case was not related to our choice of $w=3$.

\subsection{Accordion model assumptions}
\label{ssec:accordion}
We now create a generalized model for testing purposes where we have data which does not maintain regular pulses.  
This model will generate data on a window $[0,W]$.
Assume $Q_1,\cdots,Q_K$ are drawn iid from $\mathrm{unif}((1-\e)T,(1+\e)T)$.
We assume $\e \in [0,1]$ and $K>\frac{W}{(1-\e)T}$ is large enough to ensure that $\sum_{i=1}^K Q_i \geq W$.
We will generate pulses so that the length of the $i$th period is $Q_i$.

For the sake of notation, let $Q_0 = 0$.
For any $s \in [0,W]$, define 
$\sigma(s) = \max\{j \in [0,\cdots,K] \mid \sum_{i=0}^jQ_i \leq s\}$.
The reparameterization function is given by 
\begin{equation*}
\phi(s) = 
\left( 
	T\cdot \sigma(s)
\right)
	 + 
\frac{T}{Q_{\sigma(s)+1}}
\left(s - \sum_{i=0}^{\sigma(s)}Q_i\right).
\end{equation*}
Then the noisy time series is given as 
\begin{equation}
\label{eq:ModelAccordion}
X(s) = P_T^\tau(\phi(s)+\delta_x) + \delta_y
\end{equation}
where $\delta_x$ and $\delta_y$ are noise as defined in Sec.~\ref{ssec:modelAssumptions}.
See Fig.~\ref{fig:PulseTrain} for an example.
Notice that this model simplifies to Eq.~\ref{eq:ModelSimple} when $\e = 0$.
%


\section{Numerical Simulations and Robustness Analysis} 
\label{sec:robust_analysis}
This section describes the numerical simulations and shows the results of the robustness analysis. 
The basic idea is to take a standard pulse train with a small duty cycle and introduce spurious peaks at both the rising and the falling edge of the data using a uniformly distributed random variable as defined in Eq.~\ref{eq:ModelSimple}. 
This noise is found to mimic the digital ringing noticed in the laser tachometer experimental data discussed in Sec.~\ref{sec:experimental_apparatus}. 
Further, we simulate data from a signal where the spacing between the peaks is non-constant as defined in Eq.~\ref{eq:ModelAccordion}.
The signal with modulated peak separation simulates, for example, a heavy cutting process where the spindle speed varies during the cut. 
Alternatively, this spindle speed variation might be intentionally introduced to mitigate chatter and improve the cutting process \cite{Sexton1978, Yilmaz2002, Lin1990, Radulescu1997}. 
The basic parameters used in the simulation are shown in Table \ref{tab:simParams}.

\begin{table}[tbp]
\centering
\begin{tabular}{c|c|c}
Parameter & Description  & Value(s) studied \\
\hline
$\delta $& nominal duty cycle for the pulse & $5\%$ \\
$\Omega_0$ & nominal RPM & $[30, 24000]$ \\
$T$ & nominal period of the pulse in seconds & $60/\Omega_0$ \\
$\tau$ & time in on cycle &  $\delta \cdot T$\\
$\alpha$ & uniform noise in $x$ as \% of $\tau$  & $[0, 0.5]$\\
$\beta$ & uniform noise in $y$ as \% of amplitude  & 0\\
$\epsilon$ & peaks' spacing uniform noise amplitude/$T $& $[0.02, 0.65]$ \\
$n$ & number of periods to simulate & $32$ \\
$m$ & oversampling factor & $32$ \\
$N$  & number of simulation points & $N=m ( \Omega_0/60)  nT$ 
\end{tabular}
\caption{The parameters used in the numerical simulations.}
\label{tab:simParams}
\end{table}
In order to test the robustness of the persistence-based algorithm, the described numerical simulations were utilized in a number of analyses which include:
\begin{enumerate}[nosep, labelindent=0pt]
	\item Investigating the described approach as a function of digital ringing. This is controlled by the parameter $\alpha$ which specifies, as a percent of $\tau$, the interval around the rising and falling edges that will be affected by digital ringing. The digital ringing is introduced using a uniformly distributed variable defined on the interval described by $\alpha$. Therefore{}, as $\alpha$ gets larger, the digital ringing gets more severe. 
	\item Studying the performance of the method when varying the spacing between the peaks (the accordion effect). This is controlled by the parameter $\epsilon$ which represents, as a percent of $T$, the width of the interval that will be modulated by uniform noise. The larger $\epsilon$, the more likely it is for the peaks to get closer.  
	\item Analyze the results of our approach versus the well-established Fourier transform method. With Fourier the frequency is found and the RPM or the count of pulses per unit time are then deduced from the prominent frequency components.
	\item Benchmark the computational time of the persistence-based approach versus the heavily optimized Fourier transform approach.  
\end{enumerate}
All the simulations and the related analysis were performed using Python's \texttt{scipy} stack.
The numerical simulations consisted of a $100 \times 100$ uniform grid each in the ($\Omega_0$, $\alpha$) plane, and the ($\Omega_0$, $\epsilon$) plane. The simulation grids were defined using $\Omega_0$ $\in [30, 24000]$, $\alpha \in [0, 0.5]$, and $\epsilon \in [0.02, 0.65]$. For both sets of simulations, $100$ replicates were generated at each grid point. The starting seed for the random number generator used to produce the random noise was $48824$, and it was incremented by one for each subsequent iteration. For each grid point, the RPM was calculated using both persistence (Eq.~\ref{eqn:RPM}) and Fourier (Eq.~\ref{eq:RPMxFourier}). The computation time for each method was also recorded using the performance counter from Python's \texttt{time} package.

\begin{figure}[htbp]
\centering
\includegraphics[width=\textwidth, draft = False]{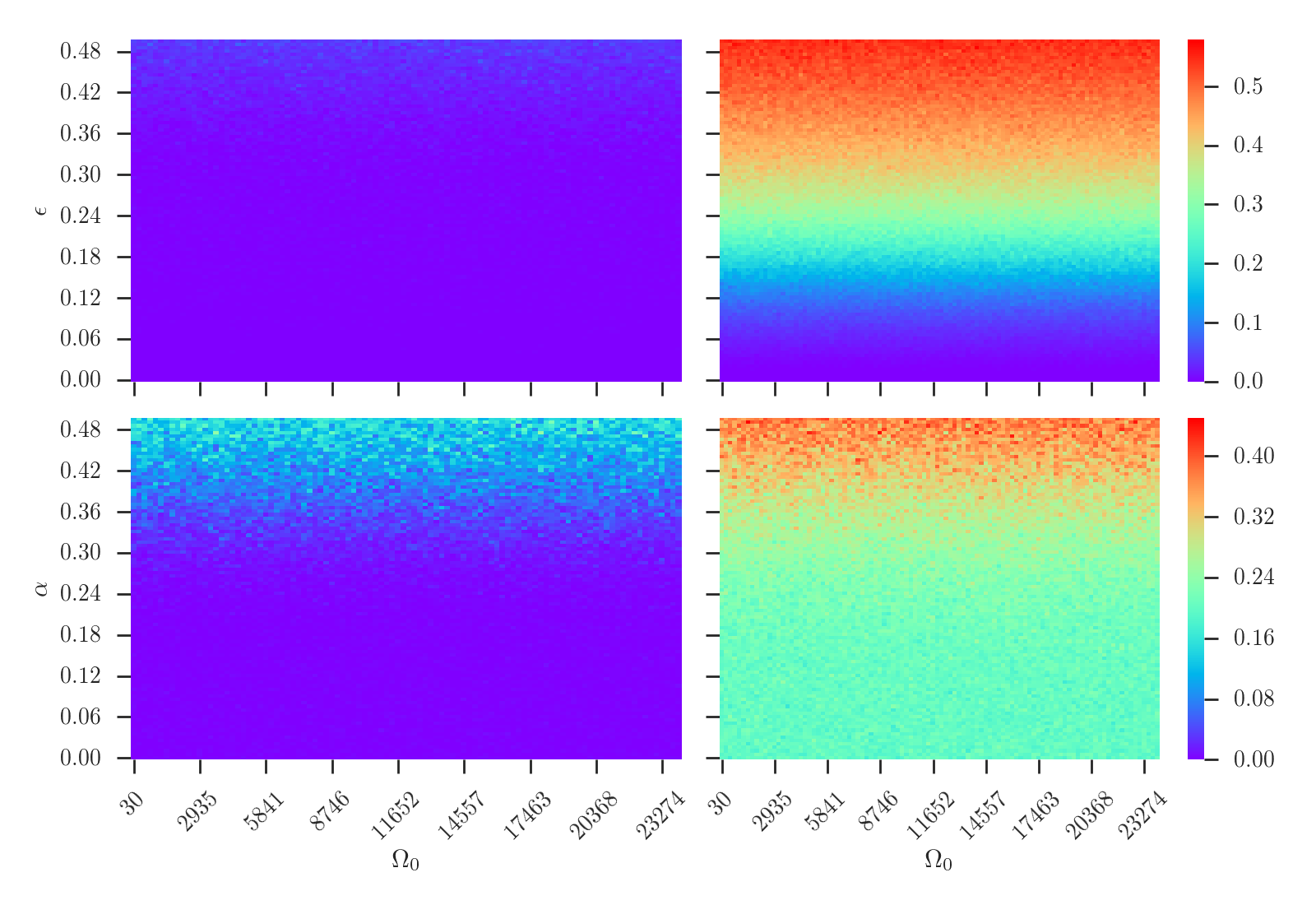}
\caption{Heat maps showing a comparison of the relative errors between the calculated RPM using persistence (left column) and Fourier (right column). The first row shows the comparison in the ($\Omega_0$, $\epsilon$) plane, while the second row compares the two algorithms in the ($\Omega_0$, $\alpha$) plane. Note that the color scale is fixed for each row to facilitate the comparisons. }
\label{fig:fullheatmap}
\end{figure}

The sample mean was used as the point estimator on each grid point.
The relative error between the output of each algorithm (Fourier, and persistence) and the nominal RPM $\Omega_0$ was computed. 
The results that show a comparison between Fourier and persistence are summarized in Fig.~\ref{fig:fullheatmap}. 
The left and right columns correspond to persistence-based and Fourier-based results, respectively. 
The first row is for varying the accordion parameter $\epsilon$, while the second row corresponds to varying the digital ringing parameter $\alpha$. The same color scale is used for both heat maps in the same row, and higher values on the color scale indicate larger relative error. 

\subsection{\texorpdfstring{$\alpha$}{alpha} versus \texorpdfstring{$\Omega_0$}{nominal RPM}}
\begin{figure}[t!]
\centering
\includegraphics[width=\textwidth, draft = False]{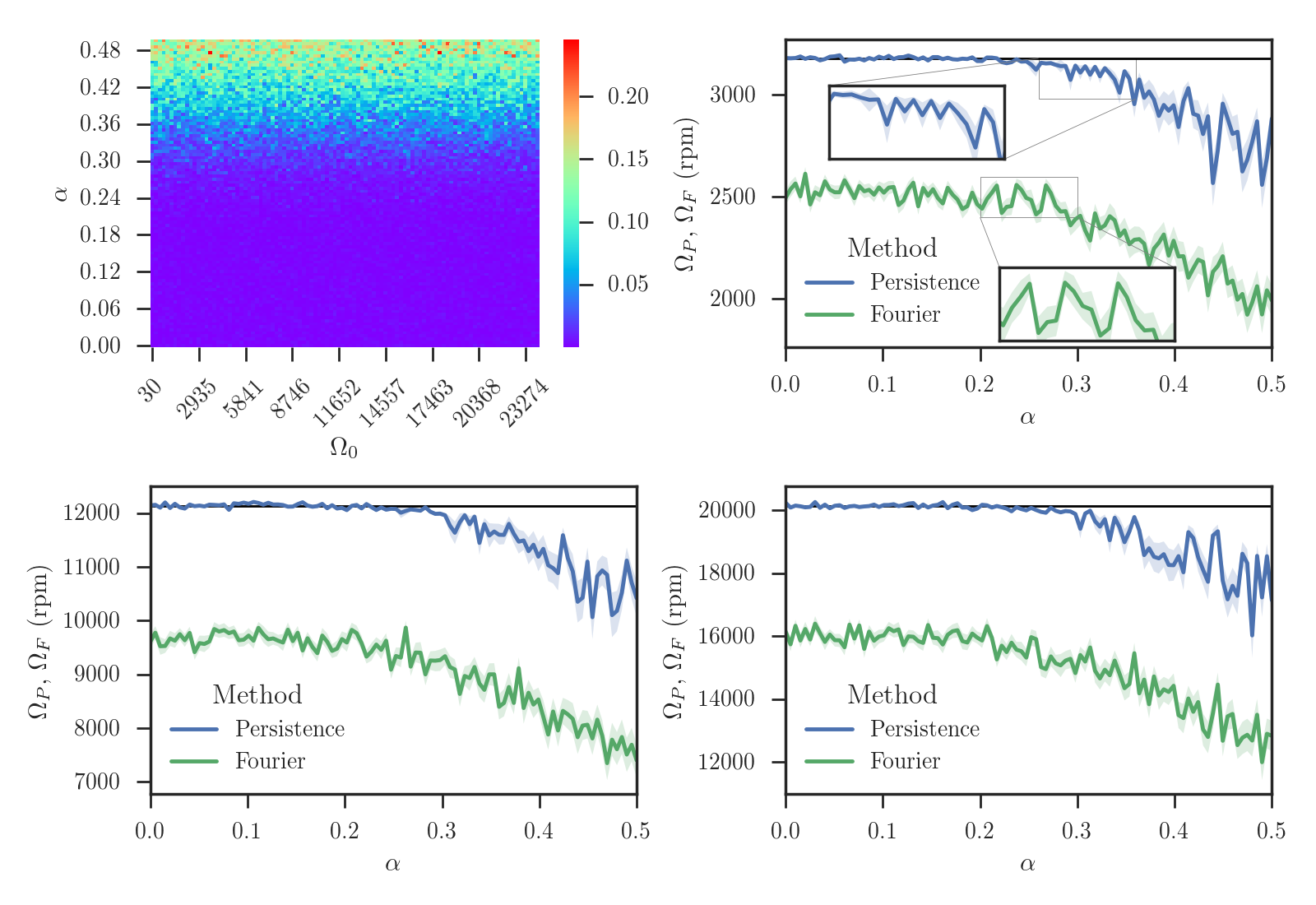}
\caption{Top left: A figure showing more details of the heat map that depicts the relative error in the persistence-based RPM as a function of $\Omega_0$ and $\alpha$. Top right, bottom left, and bottom right: The calculated RPM using persistence and Fourier algorithms for $\alpha=10\%$ and the RPM values $3178$, $12136$, and $20126$, respectively. The nominal RPM values are marked using horizontal black lines. The bands around the line represent the $68\%$ confidence bands which were computed using an empirical bootstrap distribution of the sample mean. }
\label{fig:alphaVsRpm}
\end{figure}

The left heat map in the bottom row of Fig.~\ref{fig:fullheatmap} shows that persistence performed significantly better than Fourier. Specifically, the relative error of the persistence algorithm is below $7\%$ for $\alpha < 30\%$. In contrast, the error in Fourier is as high as $24\%$ for $\alpha < 30\%$ and goes to at least $40\%$ for $\alpha > 35\%$. 
Note that in both rows of Fig.~\ref{fig:PulseTrain} the color scale is the same. A closer look at the heat map of persistence in Fig.~\ref{fig:alphaVsRpm} shows how the error in the persistence algorithm is between $15\%$ and $25\%$ but only for $\alpha > 30\%$. 
These are large values that correspond to a signal with very strong variations between the peaks. 
Note that even though we show the results for $\alpha \in [0, 0.5]$, we could deal with the case of $\alpha > 0.5$ by reversing the persistence algorithm from looking at the gaps between peaks (logic zero) to looking at logic one (high values in the two-level digital signal). 
Therefore, the user has the flexibility of adjusting the algorithm to recover good performance, as needed.

Figure~\ref{fig:epsVsRpm} shows the plot of the average RPM versus $\alpha$ calculated using both persistence and Fourier for $\epsilon=25\%$ at three nominal rpms: $3178$ (top right), $12136$ (bottom left), and $20126$ (bottom right). 
The nominal RPM is indicated on the figure using a solid, horizontal black line. 
The curves for the calculated persistence and Fourier RPM values also include the $68\%$ confidence band that can be seen in the inset panels of the top right figure. 
These confidence bands were computed using an empirical bootstrap distribution of the sample mean. 
It can be seen that for $\alpha < 20\%$, Fourier consistently underestimates all three nominal RPM values. 
Moreover, as $\alpha$ is increased beyond $20\%$, the Fourier estimate starts to decrease further below the nominal RPM. 
In contrast, persistence gives close estimates of the nominal RPM for $\alpha < 25\%$, and only tends to consistently undershoot the nominal RPM for $\alpha >25\%$. 
Therefore, the bottom rows in Figs.~\ref{fig:fullheatmap} and \ref{fig:alphaVsRpm} show that the persistence algorithm is robust to the width of the digital ringing. 

\subsection{\texorpdfstring{$\epsilon$}{epsilon} versus \texorpdfstring{$\Omega_0$}{nominal RPM}}

\begin{figure}[tb]
\centering
\includegraphics[width=\textwidth, draft = False]{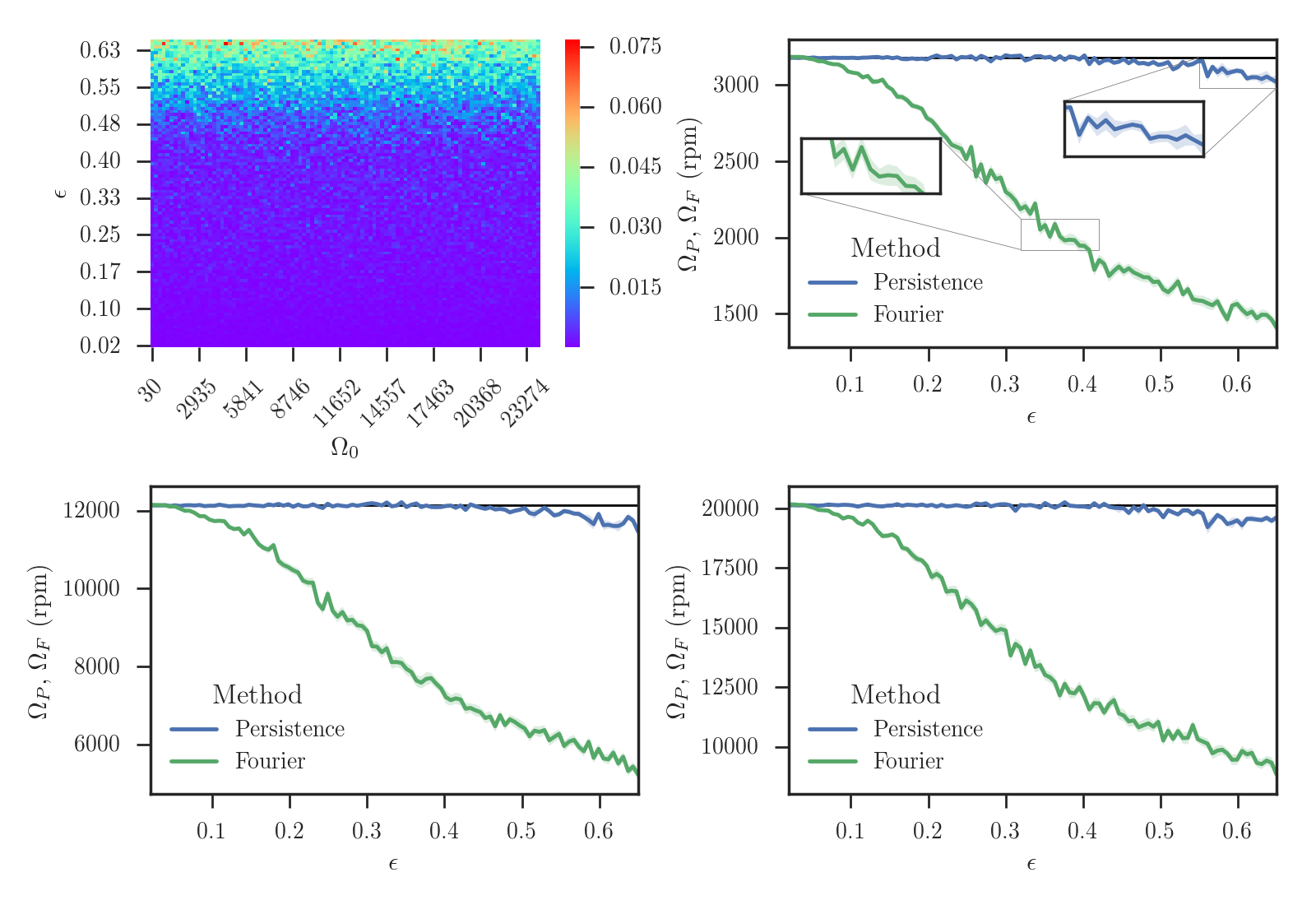}
\caption{Top left: A figure showing more details of the heat map that depicts the relative error in the persistence-based RPM as a function of the nominal RPM and $\epsilon$. Top right, bottom left, and bottom right: The calculated RPM using persistence and Fourier algorithms for $\epsilon=25\%$ and the RPM values $3178$, $12136$, and $20126$, respectively. The nominal RPM values are marked using horizontal black lines. The bands around the line represent the $68\%$ confidence bands which were computed using an empirical bootstrap distribution of the sample mean. }
\label{fig:epsVsRpm}
\end{figure}

The top row of Fig.~\ref{fig:fullheatmap} shows that for $\epsilon < 7\%$ the two algorithms produce comparable results with less than $10\%$ relative error. However, as $\epsilon$ is increased, the performance of Fourier progressively deteriorates from $10\%$ to greater than $40\%$ as evidenced by the increase in the relative error across approximately four horizontal bands with a width of $10\%$ each. These bands are roughly given by
$\epsilon \leq 15\%$,
 $18\% < \epsilon \leq 30\%$,
 $30\% < \epsilon \leq 40\%$,
and $\epsilon > 40\%$.
In contrast, the left heat map in the first row shows that persistence consistently maintains less than $10\%$ error for almost all values of $\epsilon$; however, errors start to increase for $\epsilon > 42\%$. The deterioration of the performance of the described approach is better observed from the heat map in Fig.~\ref{fig:epsVsRpm} where higher relative errors start occurring towards the upper limit of the $\epsilon$ values.

In addition to the heat map that shows the relative error as a function of $\epsilon$ and $\Omega_0$, Fig.~\ref{fig:epsVsRpm} shows the plot of the average RPM versus $\epsilon$ calculated using both persistence and Fourier for $\alpha=10\%$ at three nominal RPM values: $3178$ (top right), $12136$ (bottom left), and $20126$ (bottom right). 
The nominal RPM is indicated on the figure using a solid, horizontal black line.
The curves for the calculated persistence and Fourier rpms also includes the $68\%$ confidence band that can be seen in the inset panels of the top right figure. 
These confidence bands were computed using an empirical bootstrap distribution of the sample mean. 
We can see that for both algorithms the confidence bands are narrow. 
Further, it can be seen that for $\epsilon < 5\%$ both algorithms are close to the nominal value for all three RPM choices. 
However, for $\epsilon > 5\%$, the Fourier algorithm quickly diverges away from the nominal RPM while persistence remains close to the nominal value. 
As $\epsilon$ is increased to $45\%$, we start to see persistence undershooting the nominal value which indicates that at high values of $\epsilon$, persistence undercounts the true peaks. 
Therefore, the top rows in Figs.~\ref{fig:fullheatmap} and \ref{fig:epsVsRpm} show that the persistence algorithm is robust to the variation in the spacing between the peaks.

\subsection{Runtime comparison} \label{sec:runtime}
\begin{figure}[t!bp]
\centering
\includegraphics[width=0.75\textwidth, draft = False]{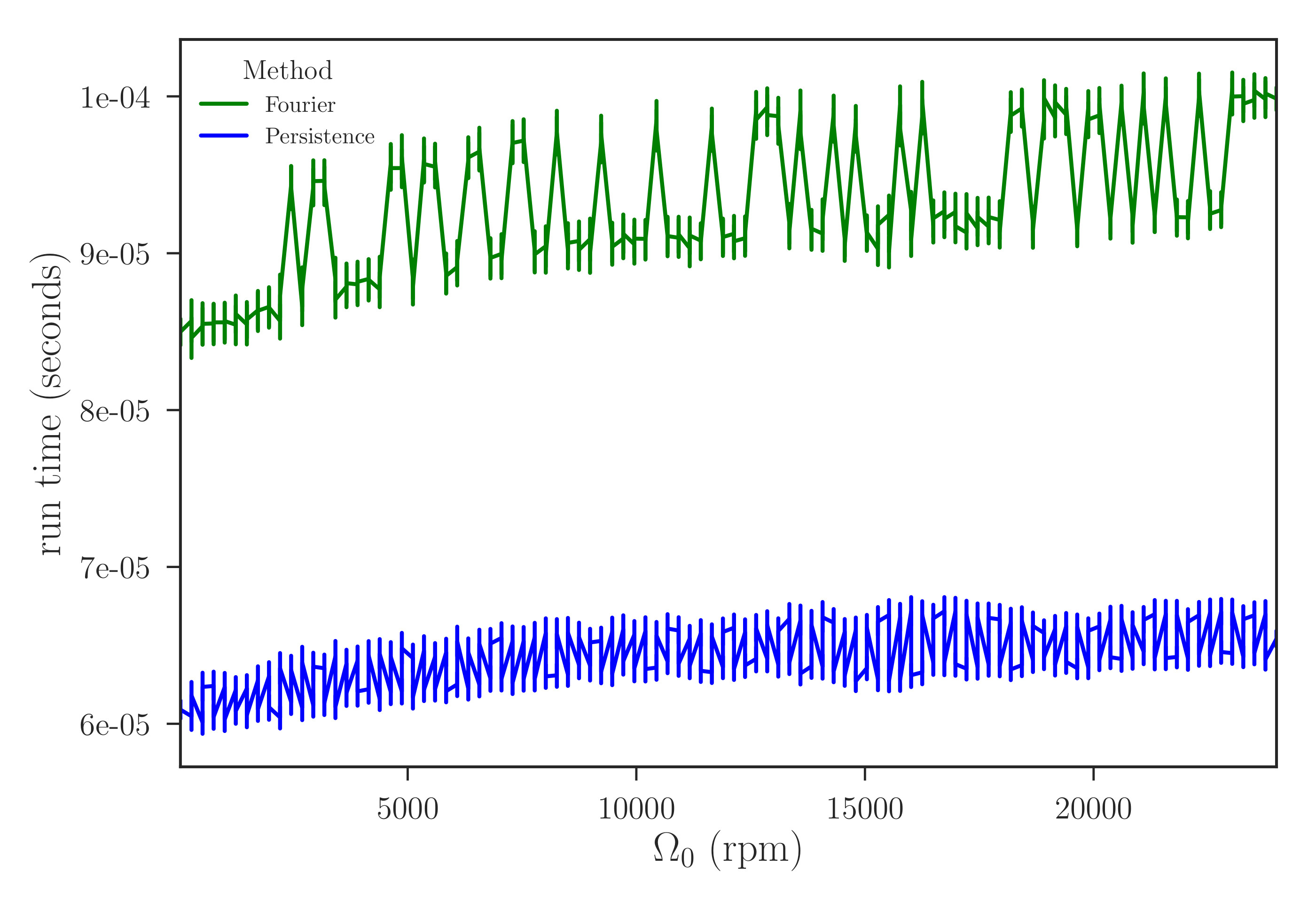}
\caption{Average runtime for all trials as a function of the nominal RPM, i.e., the signal length.}
\label{fig:runtime}
\end{figure}
In terms of theoretical runtime, the algorithm for computing $\Omega_P$ is simple, requiring only a sequence of pairwise subtractions and sortings.  
Since these take $O(n)$ and $O(n\log n)$ respectively, the theoretical worst case runtime for the persistence based algorithm is $O(n\log n)$.  
Meanwhile, the FFT algorithm computes the Fourier transform in $O(n\log n)$ as well, so the worst case analysis of the two methods is the same.

Runtime benchmarks for Fourier and persistence algorithms were measured using the performance counter from Python's \texttt{time} package. 
The machine used for the simulations is a standard desktop running Ubuntu 16.04 with 32GB ram and a 3.6GHz Intel i7 processor. 
A total of $200$ runs for each algorithm was performed at each value of the nominal RPM. 
The nominal RPM is kept as a variable of runtime because it controls the length of the simulated pulse (see Table \ref{tab:simParams}), and consequently, the runtime for each algorithm. The different runs were then averaged for each algorithm, and the results are plotted in Fig.~\ref{fig:runtime}. 
The line in the figure describes the average while the bands represent the $68\%$ confidence interval obtained by bootstrap sampling of the replicates. Notice the small $y$-axis scale in Fig.~\ref{fig:runtime} indicating that both algorithms run relatively fast. However, the figure shows that persistence runs slightly faster than Fourier, and that the runtime is approximately constant across the RPM range. 
In contrast, the runtime for Fourier varies more than its persistence counterpart for the different values of the nominal RPM. 
%

\section{Experimental verification} 
\label{sec:experimental_apparatus}
\begin{figure}[tbp]
	\centering
	\includegraphics[width=0.45\textwidth]{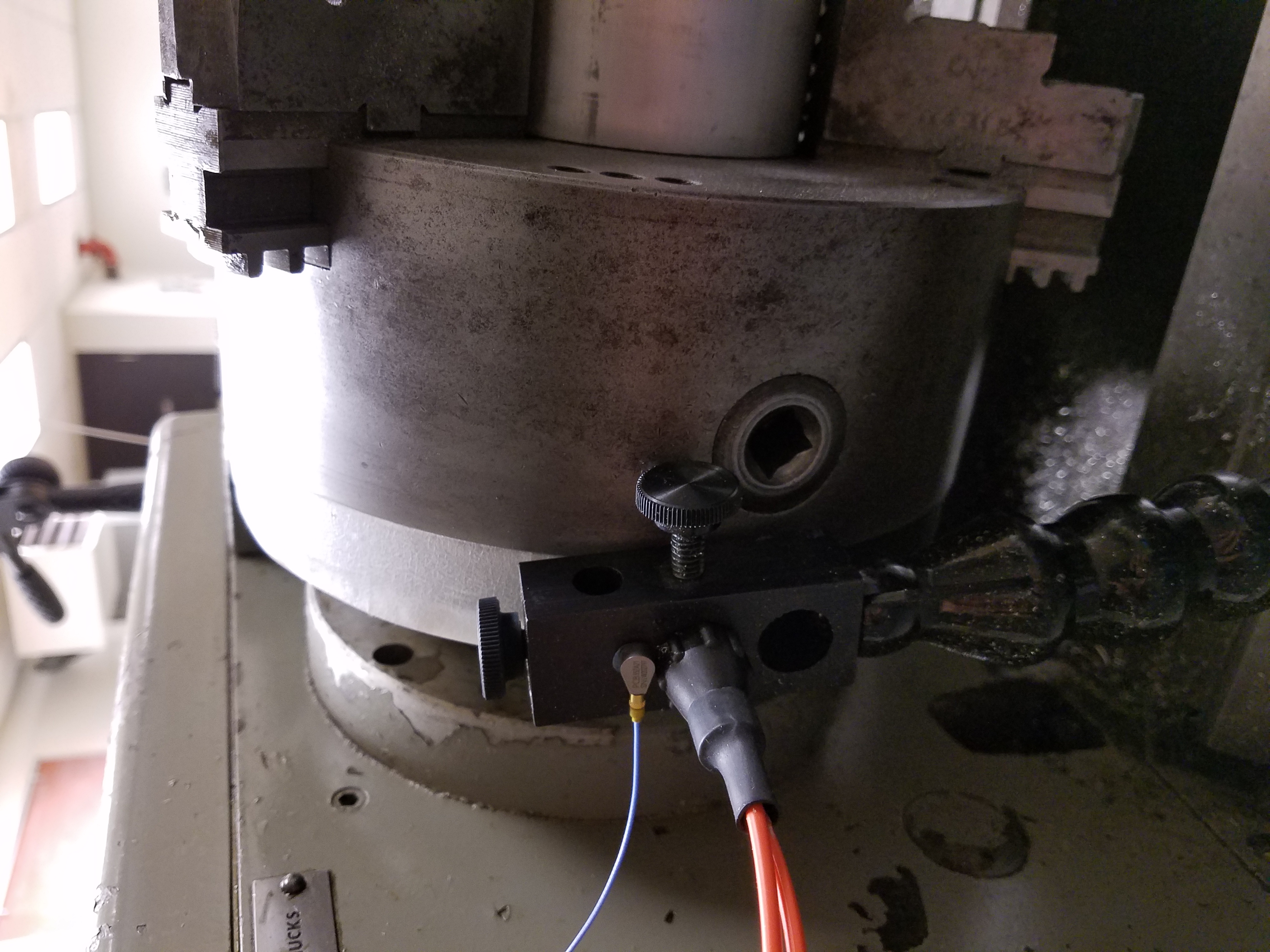}
	\hfill
	\includegraphics[width=0.45\textwidth]{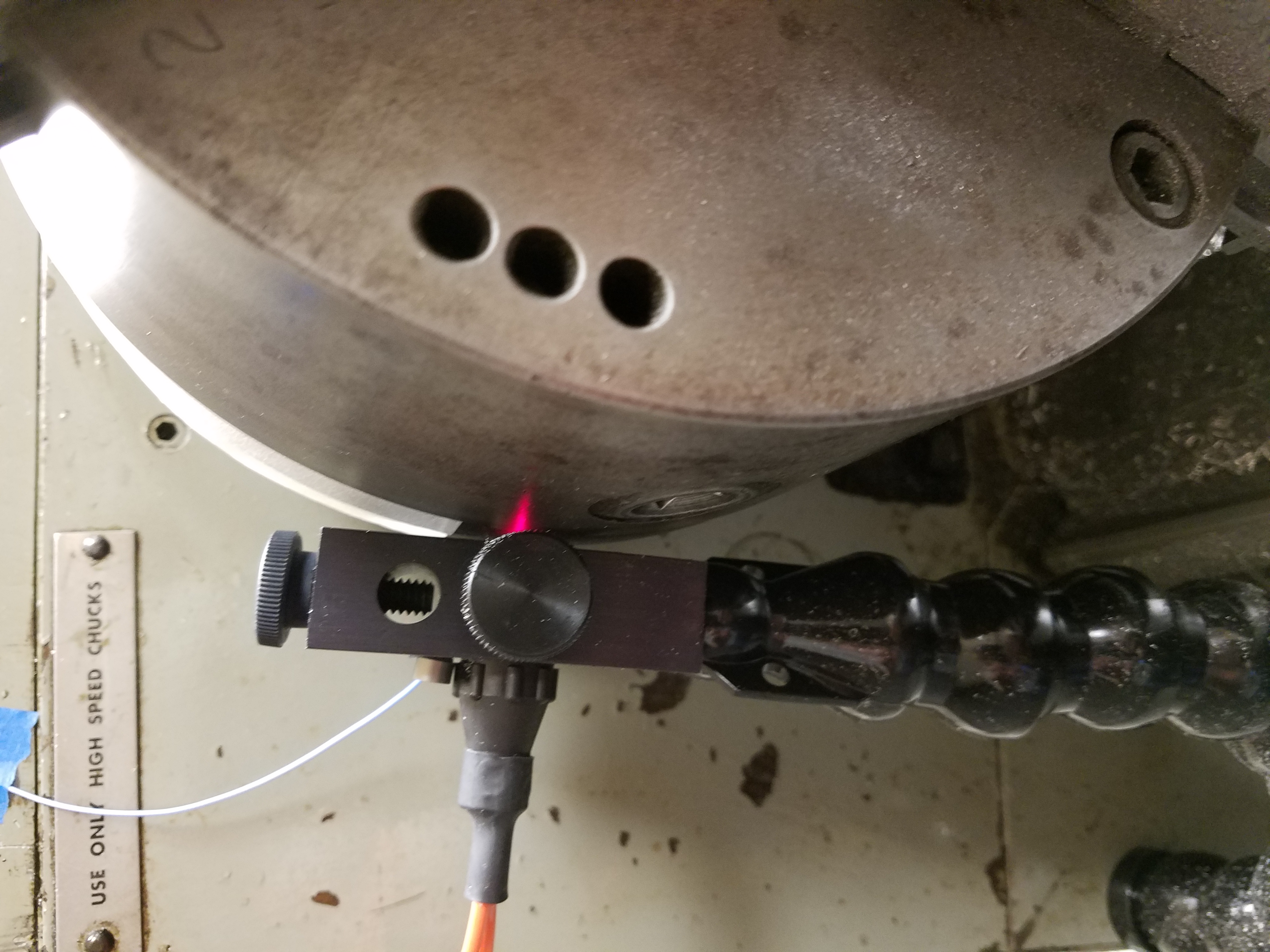}
	\caption{The experimental apparatus consists of a laser tachometer mounted against a rotating spindle. The left picture shows the back view while the right picture shows the side view. Half of the spindle is fitted with white tape whereas the other half is covered with black tape.}
	\label{fig:lasertachExp}
\end{figure}
The experimental apparatus is shown in Fig.~\ref{fig:lasertachExp}. It consists of a Terhahertz Technologies LT-880 Laser Tachometer mounted against the spindle of a Clausing-Gamet 33 Centimeters (13 inch) engine lathe. The laser beam is emitted onto a point along the circumference of the spindle. Half of the spindle is covered with white tape, while the other half is covered with black tape. When the emitted light beam is reflected back to the tachometer's receiving lens, a 5 volt TTL pulse is registered. These pulses are captured by an NI USB-6366 data acquisition box using Matlab. The data is collected using an analog channel to emulate a common practice in industry where sensory inputs in machining processes (including those of laser tachometers) are all collected using analog channels of the same data acquisition device. The signal to noise ratio for laser tachometers is large and it is easy to recover a digital signal by hard-thresholding. In this study, additive normal noise was removed by setting the threshold at 2.5 volts. Thus, any signal above the threshold was set to 1, while signals below the threshold were set to zero.

The experiment was performed using several nominal spindle speeds, which are shown in Table \ref{tab:nominalrpm}. Three trials were performed for each speed and data collection started after the initial spindle ramp up ended. The data was sampled at 80 kHz and the duration of the data collection was chosen such that for each trial the recorded time series captured $30$--$32$ true peaks in the tachometer's pulse train; again, see Table \ref{tab:nominalrpm}. 

\begin{figure}[htbp]
	\centering
	\includegraphics[width=0.85\textwidth, draft = False]{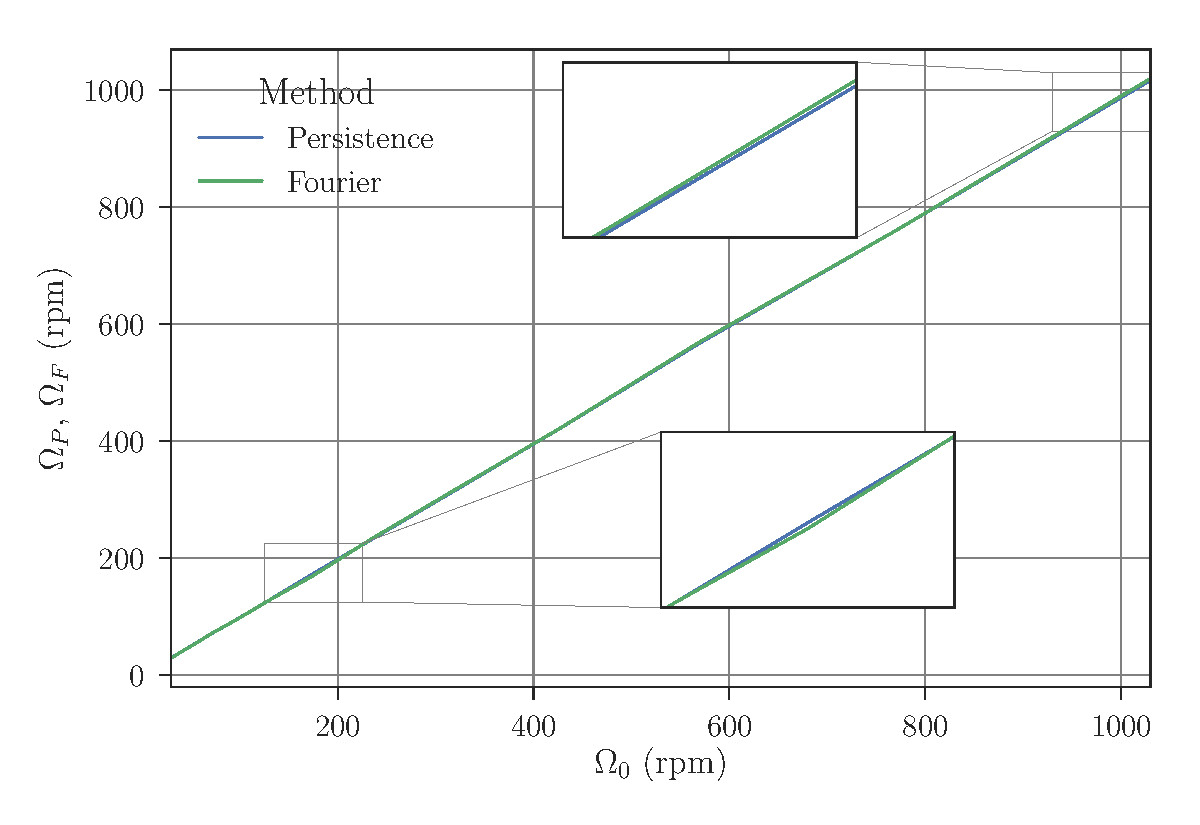}
	\caption{Comparison of $\Omega_P$, $\Omega_F$, and the nominal spindle speeds using experimental data. The inset panels shows that both $\Omega_P$ and $\Omega_F$ are close to each other and to the nominal speed even at the sections where they deviate the most. The figure shows that both $\Omega_P$ and $\Omega_F$ give a speed that is lower than its nominal value for $\Omega_0 > 800$ RPM. This is attributed to the chuck slip effects at the higher range of speeds. }
	\label{fig:experimental_rpm}
\end{figure}

The reason for choosing a sampling rate of 80 kHz is twofold. 
First, it is now becoming standard to use digital signal processing (DSP) for low/high/band-pass filtering instead of in-line analog filters. 
A conservative rule of thumb in DSP is to oversample by a factor of 16, digitally filter the signal, and then downsample to the frequency range of interest. 
Since many signals in cutting processes are collected to study or prevent chatter, which can occur at frequencies of a few thousand hertz, we wanted to make sure that our results apply to cases where by the time all the signals being collected are downsampled by a factor of 16 we can recover a noise-free signal within the range of 5 kHz; so $16 \times 5 = 80$ kHz. 
The second reason is that the frequency roll-off of the laser tachometer we are using is 40 kHz. 
Therefore, using 80 kHz ensures that we are utilizing the full dynamic range of the sensor. 
This eliminates the possibility of attributing the digital ringing we are observing in the signal to insufficient sampling. 
Nevertheless, although we are reporting the results when sampling at 80 kHz, we tried other, lower oversampling ratios, e.g., by a factor of 4, and the observed behavior was the same.

Figure \ref{fig:experimental_rpm} shows the results of calculating $\Omega_{\text{P}}$ and $\Omega_{\text{F}}$ using the collected signals. It can be seen that Fourier and persistence results are almost indistinguishable. Further, close examination of the the two inset panels in Fig.~\ref{fig:experimental_rpm}  show that the two curves can cross at several points and trade places. Nevertheless, they remain close and parallel for all practical reasons. A grid is superimposed on the figure to show the expected points on a curve that represents perfect RPM calculation versus $\Omega_0$, the nominal RPM set on the machine. It can be seen from the figure that both $\Omega_{\text{P}}$ and $\Omega_{\text{F}}$ match $\Omega_0$ for $\Omega_0 < 800$ RPM. However, for $\Omega_0 \geq 800$, both $\Omega_{\text{P}}$ and $\Omega_{\text{F}}$ undershoot $\Omega_0$. This was expected since the machine for these spindle speeds was operating at its high range and therefore, spindle slipping and inaccuracies in the $\Omega_0$ were expected. The registered $\Omega_{\text{P}}$ and $\Omega_{\text{F}}$, which are considered more accurate than the nominal value set on the machine, confirm this expectation. 

\begin{table}[htbp]
\centering
\begin{tabular}{|ccccccccccccc|}
\hline 
\multicolumn{13}{|c|}{Nominal RPM} \\
30 & 40 & 54 & 72 & 98 & 130 & 175 & 235 & 320 & 425 & 570 & 770 & 1030 \\  
\hline
62 & 50 & 35 & 30 & 20 & 15  & 12  & 9   & 7   & 5   & 4   & 3   & 3   \\
\multicolumn{13}{|c|}{Duration of data record (seconds)} \\
\hline 
\end{tabular}
\caption{The nominal spindle speeds (top block) and the corresponding duration of the data collection for each trial (bottom block). Three records were collected at each nominal RPM. }
\label{tab:nominalrpm}
\end{table}

%

\section{Conclusions and Discussion}
\label{sec:Conclusions}
This paper described a new approach for step detection using tools from topological data analysis, specifically, 0-dimensional persistent homology.
This viewpoint allows us to provide guarantees via Thm.~\ref{thm:MainResult} for counting pulses even in the presence of digital ringing and stochasticity in period length.
Specifically, the approach involves computing the persistence diagram for the set of times when the time series is above a threshold, and using the widest split in this diagram to determine how many true pulses were seen.
The method is further extended to be able to give an estimation of RPM.

The described approach was verified using numerical and experimental studies. 
Specifically, simulated pulse trains with small duty cycles were generated using two independent uniform noise components: one that varied the spacing between the pulses $\epsilon$, and one that introduced spurious peaks near the rising and falling edges of the signal $\alpha$. 
The $\alpha$ noise component simulates the digital ringing observed in the experimental data. 
The top row of Fig.~\ref{fig:fullheatmap} shows that while both Fourier and persistence methods are resilient to $\epsilon < 7\%$, persistence is better suited for higher $\epsilon$ values. 
Specifically, for the Fourier-based approach the relative error progressively grows to about $40\%$ across four horizontal bands of $\epsilon$ intervals while the relative error of the persistence-based approach remains below $10\%$.

The bottom row of Fig.~\ref{fig:fullheatmap} shows that the persistence based approach gives results with a relative error below $7\%$ for $\alpha < 30\%$. 
The Fourier-based approach gives errors as high as $24\%$ for the same $\alpha$ range.  
Further, the figure shows that while the persistence-based approach gives a relative error between $15\%$ and $25\%$ for $\alpha > 30\%$, the error in the Fourier-based approach is at least $40\%$. 
Beyond $42\%$, the described approach can regain its accuracy by detecting high logic between valleys as opposed to detecting low logic between peaks. 

A close examination of sample trajectories of $\epsilon$ versus $\Omega_{\text{P}}$ and $\Omega_{\text{F}}$ in Fig.~\ref{fig:epsVsRpm} shows that both methods have tight error bounds. 
However, the persistence-based method remains close to $\Omega_0$ while $\Omega_{\text{F}}$ deviates rapidly from $\Omega_0$ for $\epsilon > 5\%$.

Similarly, Fig.~\ref{fig:alphaVsRpm} shows that persistence remains close to $\Omega_0$ with noticeable deviations from $\Omega_0$ starting at $\alpha > 30\%$. 
The same figure shows that Fourier-based approach deviates from $\Omega_0$ over the whole $ 0.02 < \alpha < 0.5$ range. 

Computationally, persistence and Fourier have the same theoretical runtime as shown in Section \ref{sec:runtime}. 
This is confirmed by the benchmark shown in Fig.~\ref{fig:runtime} where it is seen that the average runtime of both algorithms is small although persistence is consistently smaller. 
Further, the figure shows larger variations in the runtime of Fourier in comparison to persistence. 
The $68\%$ confidence band around the two curves is extremely small, which indicates high confidence in the mean as an estimator of the true runtime.

The experimental investigations depicted in Fig.~\ref{fig:lasertachExp} further established the reliability of the described approach.
Specifically, Fig.~\ref{fig:experimental_rpm} shows that both persistence-based and Fourier-based calculated RPMs are very close. 
For the specific lathe used in the experiment, both methods gave spindle speeds close to the nominal value for $\Omega_0 < 800$ RPM. 
However, for $\Omega_0 > 800$ RPM, both $\Omega_{\text{P}}$ and $\Omega_{\text{F}}$ yielded RPM values smaller than the nominal. 
This discrepancy at higher speeds is attributed to operating the machine at its high range where slipping in the spindle drive can cause the actual RPM to be smaller than the set RPM.

One caveat we mention here is that it is important to pick a span of time that contains a reasonable number of true peaks. 
If the picked time span is too long, then the resulting distribution of the gaps (or the peaks if the algorithm is switched around) may fill up the histogram of the points in the persistence diagram (see Fig.~\ref{fig:point_cloud_cluster_hist}) which can make distinguishing the true peaks difficult or impossible.

\subsection{Higher dimensional extensions}
\label{ssec:higherDim}
In essence, we have taken a difficult problem --- clustering points in $\R^d$ --- to a rather simplified setting --- $d=1$.
This restriction allows for very fast algorithms, particularly owing to the fact that one knows the minimal spanning tree in advance. 
Still, algorithms for computing 0-dimensional persistent homology for points in higher dimensions are still quite fast: an implementation using a modified union find algorithm runs in $O(n\alpha(n))$ where $n$ is the number of points and  $\alpha$ is the  notoriously slow growing inverse Ackermann function.

\begin{figure}[tb]
	\centering
	\includegraphics[draft =False, height = 2in]{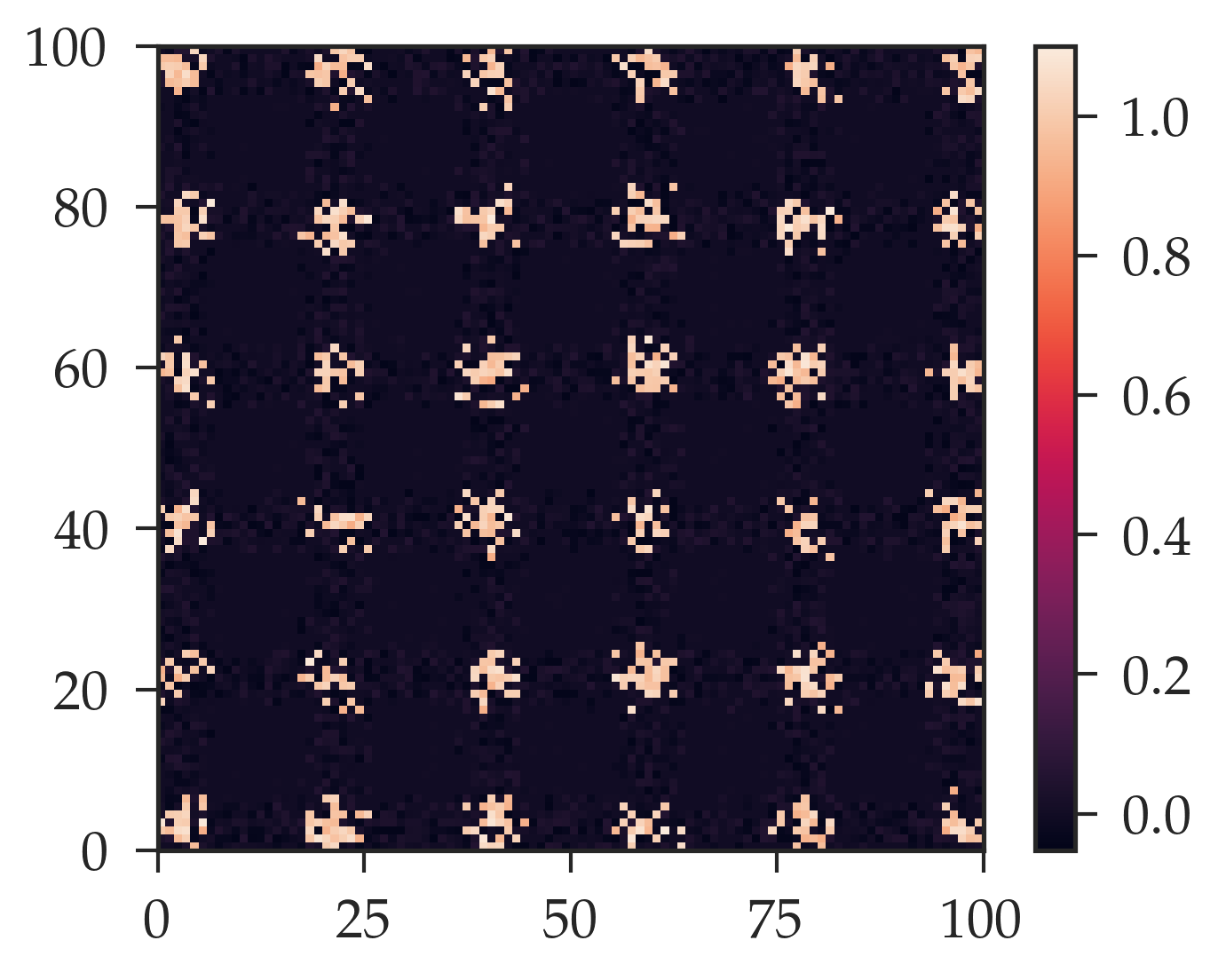}
	\includegraphics[draft = False, height = 2in]{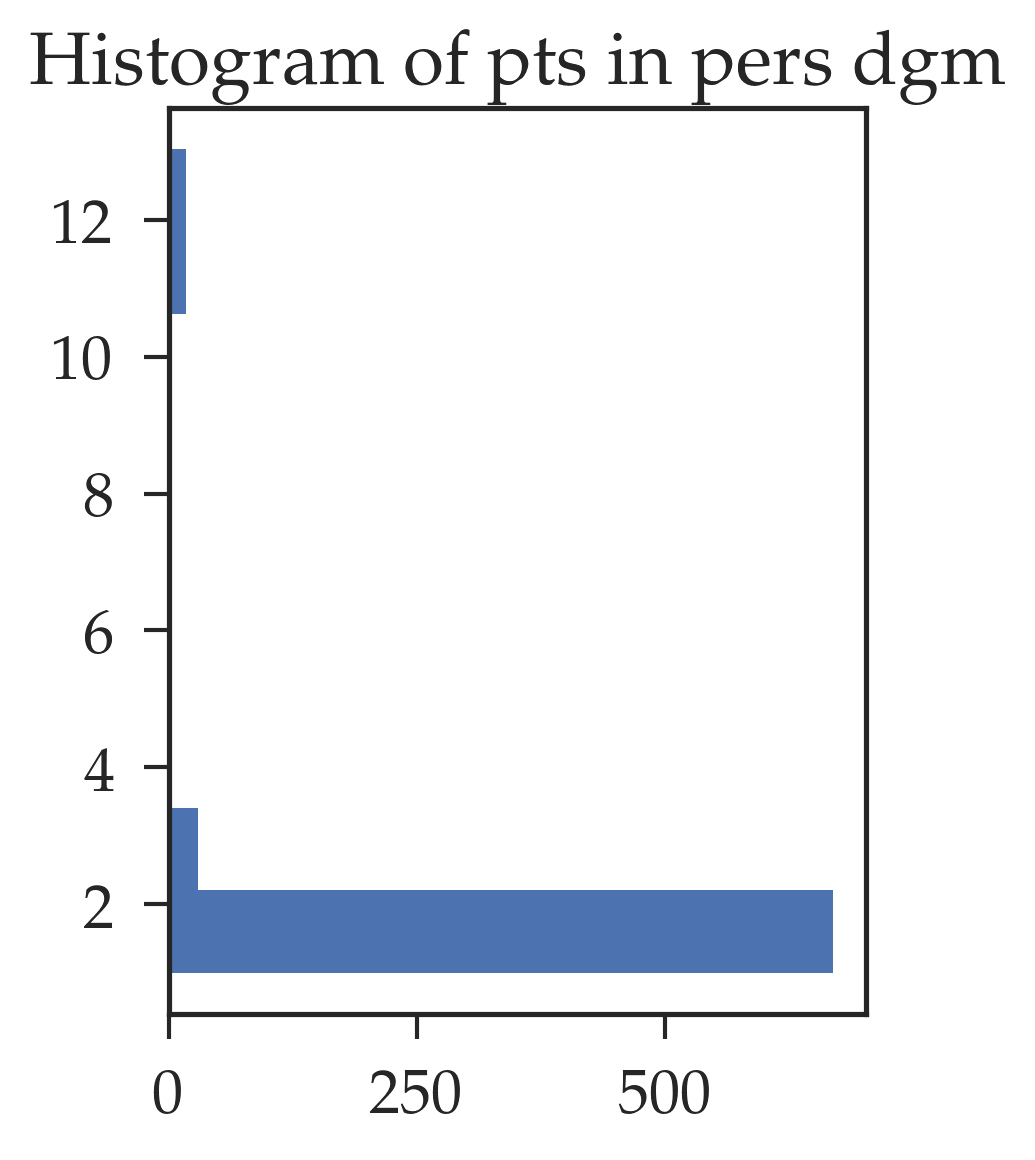}
	\caption{An example of a higher dimensional analogue of the given method for pulse counting.  The image is a product of two realizations of Eqn.~\ref{eq:ModelSimple} and the histogram shows the resulting 0-dimensional persistence diagram.  The fact that there is a distinct split between the points in the diagram means that we can still use the high-persistence points to count the number of clusters seen at left.}
	\label{fig:MultiD}
\end{figure}

One could, however, imagine higher dimensional signals (i.e., images) where there are sufficient guarantees on the pulse rate and noise inputs so as to create higher dimensional analogues of Thm.~\ref{thm:MainResult}.  
Consider, for example, the image of Fig.~\ref{fig:MultiD} generated by $Y(s,t) = X_1(s)\cdot X_2(t)$ for two realizations of Eqn.~\ref{eq:ModelSimple}.
The 0-dimensional persistence diagram histogram is shown to the right, and, as with the previous data shown in this paper, there is a clear distinction between the points.  
Since there are 35 points above a threshold of 8 in the histogram, we can conclude that there are  36 clusters in the image, which can be confirmed by inspection.
Thus, we hypothesize that this method is quite useful in higher dimensional applications, although such an application is still an open question for future research.

\blfootnote{The data files, code, and experimental parameters can be found at the gitlab repository \url{https://gitlab.msu.edu/TSAwithTDA/TDA-for-true-step-detection-in-PWC-signals}.}



\blfootnote{The authors acknowledge the help of David Petrushenko in setting up the lathe for the experiments.}



\bibliography{LaserTach}

\end{document}